\documentclass{IEEEtran}
\usepackage{cite}
\usepackage{amsmath,amssymb,amsfonts}
\usepackage{algorithmic}
\usepackage{graphicx}
\usepackage{epstopdf}
\usepackage{textcomp}
\usepackage{multirow}
\usepackage{booktabs}
\usepackage{amsmath}
\usepackage{float}
\usepackage{stfloats}
\usepackage{threeparttable}
\usepackage{booktabs}
\usepackage{tablefootnote}
\usepackage{xcolor}
\allowdisplaybreaks[4]
\def\BibTeX{{\rm B\kern-.05em{\sc i\kern-.025em b}\kern-.08em
    T\kern-.1667em\lower.7ex\hbox{E}\kern-.125emX}}
\begin{document}
\title{A SBP-SAT FDTD Subgridding Method Using Staggered Yee’s Grids Without Modifying Field Components}
\author{Yuhui Wang,~\IEEEmembership{Graduate Student Member,~IEEE,} Yu Cheng, Xiang-Hua Wang, \\Shunchuan Yang, \IEEEmembership{Member,~IEEE}, and Zhizhang Chen, \IEEEmembership{Fellow, IEEE}
\thanks{Manuscript received xxx; revised xxx.}
\thanks{This work was supported in part by the National Natural Science Foundation of China under Grant 62141405, 62071125, in part by Defense Industrial Technology Development Program under Grant JCKY2019601C005, in part by Pre-Research Project under Grant J2019-VIII-0009-0170 and Fundamental Research Funds for the Central Universities.}
\thanks{Yuhui Wang and Shunchuan Yang are with Research Institute for Frontier Science and School of Electronic and Information Engineering, Beihang University, Beijing, China (e-mail: yhwang\_0420@buaa.edu.cn, scyang@buaa.edu.cn).}
\thanks{Yu Cheng is with the School of Electronic and Information Engineering, Beihang University, Beijing, 100083, China (e-mail: yucheng@buaa.edu.cn).}
\thanks{Xiang-Hua Wang is with the School of Science, Tianjin University of Technology and Education, Tianjin, 300222, China (e-mail: xhwang199@outlook.com.)}
\thanks{Zhizhang Chen is currently with the College of Physics and Information Engineering, Fuzhou University, Fuzhou, Fujian, China, on leave from the Department of Electrical and Computer Engineering, Dalhousie University, Halifax, Nova Scotia, Canada B3H 4R2  (email: zz.chen@ieee.org). }
}

\maketitle

\begin{abstract}
A summation-by-parts simultaneous approximation term (SBP-SAT) finite-difference time-domain (FDTD) subgridding method is proposed to model geometrically fine structures in this paper. Compared with our previous work, the proposed SBP-SAT FDTD method uses the staggered Yee’s grid without adding or modifying any field components through field extrapolation on the boundaries to make the discrete operators satisfy the SBP property. The accuracy of extrapolation keeps consistency with that of the second-order finite-difference scheme near the boundaries. In addition, the SATs are used to weakly enforce the tangential boundary conditions between multiple mesh blocks with different mesh sizes. With carefully designed interpolation matrices and selected free parameters of the SATs, no dissipation occurs in the whole computational domain. Therefore, its long-time stability is theoretically guaranteed. Three numerical examples are carried out to validate its effectiveness. Results show that the proposed SBP-SAT FDTD subgridding method is stable, accurate, efficient, and easy to implement based on existing FDTD codes with only a few modifications.
\end{abstract}

\begin{IEEEkeywords}
Finite-difference time-domain (FDTD), subgridding, summation-by-parts simultaneous approximation term (SBP-SAT), stability

\end{IEEEkeywords}

\section{Introduction}
\label{sec:introduction}
\IEEEPARstart{T}{he} finite-difference time-domain (FDTD) method is widely used in the practical engineering applications, such as the designs of microwave device \cite{2}, fiber \cite{3}, and antenna \cite{4}, where their properties of wideband frequency response can be obtained through a single simulation \cite{5}. However, since the structural orthogonal hexagonal cells consist of the Yee’s grid, staircase errors can severely degenerate the accuracy. 

The subgridding technique is an effective approach to mitigate this issue, which can significantly improve the accuracy without compromising the efficiency, since fine meshes are only used to model geometrically fine structures and relatively coarse meshes are used in other domains. Many efforts have been made in the last few decades, such as the time- and space- separation subgridding method \cite{1-23}, the subgridding method based on wave equations \cite{1-8,1-13,1-14}, the Huygens subgridding method \cite{1-15}, the lumped element based subgridding method \cite{1-17,1-18}, the hybrid subgridding methods \cite{1-9,1-10,1-11,1-12,1-16,1-24}, the adaptive mesh refinement FDTD method \cite{1-20}.

However, one main challenge for them in the practical engineering applications is that the long-time stability cannot be always guaranteed since their analytical proofs are hard to be given. There are many efforts made to address this issue. In \cite{11}, a conditionally stable FDTD subgridding method for the unsymmetric discrete system was proposed to improve the accuracy and efficiency. Furthermore, another two symmetric and stable subgridding algorithms were proposed in \cite{12,13}. Recently, a dissipation theory was proposed to prove the stability of the FDTD methods with applications to subgridding algorithms \cite{14,15}  in two- and three-dimensional space. 

The SBP-SAT techniques were originally developed to solve the partial differential equations (PDEs), such as the Euler and Navier-Stokes equations \cite{16,17}, wave equations \cite{18,19}, through the provably stable finite-difference methods in the long-time simulations. The SBP operator can make the energy estimate in the computational domain fully determined by the boundary values, and then the boundary conditions are weakly enforced through the SATs. The long-time stability of the SBP-SAT finite difference methods is theoretically guaranteed \cite{20,21}. In the computational electromagnetics society, there are only a few literatures upon their applications in solving the Maxwell’s equations. In \cite{22}, a two-dimensional SBP operator with collocated grids and a one-dimensional SBP operator with staggered grids were proposed for the Maxwell-Duffing models. To further extend the capability in the Maxwell’s equations, a provably stable SAT-SBP FDTD method with the staggered grid is proposed, and the interpolation matrices for stable coupling multiple mesh blocks are also investigated in \cite{1}. 

In this paper, we proposed another provably stable SBP-SAT FDTD subgridding method to solve the two-dimensional Maxwell’s equations under the transverse magnetic (TM) assumption. In the proposed SBP-SAT FDTD method, the Yee’s grids without any modifications are used. To make the discrete operators satisfy the SBP property, the magnetic fields on the boundaries are extrapolated through the carefully designed projection operator, which can keep the accuracy of magnetic fields consistency with the finite-difference operator near the boundaries. Therefore, the SBP property can be met in the proposed SBP-SAT FDTD method even if no additional field components are added on the boundaries \cite{23}. When the multiple mesh blocks with different mesh sizes are used, the SATs are used to weakly enforce the boundary condition. Then, a theoretically stably SBP-SAT FDTD method with multiple mesh blocks of different mesh sizes is developed in the long-time simulations. 

Compared with the work in \cite{22}, the collocated grid is used for simulations, which shows significant different from the traditional FDTD method with the staggered Yee’s grids. As mentioned before, the SBP-SAT FDTD method with the staggered grid was proposed in \cite{1} to further extend its capability in solving the electromagnetic problems. However, four additional electric nodes are added at the corners of the computational domain, and several other electric nodes are added in the middle of edges of boundary cells. Compared with the existing and our previous works in \cite{1}, our work in this paper advances in the following three aspects. 

\begin{enumerate}
\item An SBP-SAT FDTD method with the staggered Yee’s grid without adding or modifying field components is proposed to solve the Maxwell equations in the two-dimensional space. By carefully defining the projection operator near the boundaries, the SBP property of the discrete differential operator can be satisfied. Therefore, the proposed SBP-SAT FDTD method is stable in the long-time simulations.  
\item The second-order finite-difference scheme is used and detailed operators are analytically derived to satisfy the SBP property. The energy in the whole computational domain is fully determined by field values on the boundaries.
\item Different grid blocks can be stably coupled by using the SATs, and different mesh size ratio between adjacent grid blocks is allowed due to careful designed interpolation matrices and careful selection of the free parameters in the SATs. The stability of the SBP-SAT FDTD subgridding method is theoretically guaranteed.
\end{enumerate}

The remaining of this paper is organized as follows. In Section II, the symbolic notations are defined, and then the detailed derivation of the proposed SBP-SAT method is presented. In Section III, stable coupling with multiple mesh blocks is presented. In Section IV, three numerical examples are carried out to validate the proposed SBP-SAT FDTD method. Finally, we draw some conclusions in Section V.

\section{The Proposed Fdtd Method With The SBP-SAT Technique}
\subsection{Field Distribution in the Mesh Block and Notations}
Since the Yee’s grid is used in the proposed SBP-SAT FDTD method, the field locations in the grid are exactly the same as those in the traditional FDTD method, as shown in Fig. 1. ${\bf{E}}$ and ${\bf{H}}$ nodes are interlaced with each other. ${\bf{E}}$ nodes are located at the corners of each cell and ${\bf{H}}$ nodes are placed at the middle of the edges of each cell. Compared with our previous work in \cite{1}, where additional electric nodes are added on the boundaries to make the discrete operators satisfy the SBP property, field values on the boundaries are extrapolated when necessary.
 
To make the derivation clear, the notations, which will be frequently used in the following derivation, are defined as follows.
\begin{enumerate}
\item A hollow character denotes a matrix, e.g., ${\mathbb{D}}$, and a character in bold is a column vector, e.g., ${\bf{Z}}$.
\item $\otimes$ denotes the Kronecker product of two matrices, e.g., ${{\mathbb{C}} = {\mathbb{A}} \otimes {\mathbb{B}}}$, in which the dimensions of ${\mathbb{C}}$, ${\mathbb{A}}$ and ${\mathbb{B}}$ are $N_{x 1 }N_{x 2 } \times N_{y 1 }N_{y 2 }$, $N_{x 1 } \times N_{y 1 }$, $N_{x 2 } \times N_{y 2 }$, respectively.
\end{enumerate}

\begin{figure} 	
	\centerline{\includegraphics[scale=1.2]{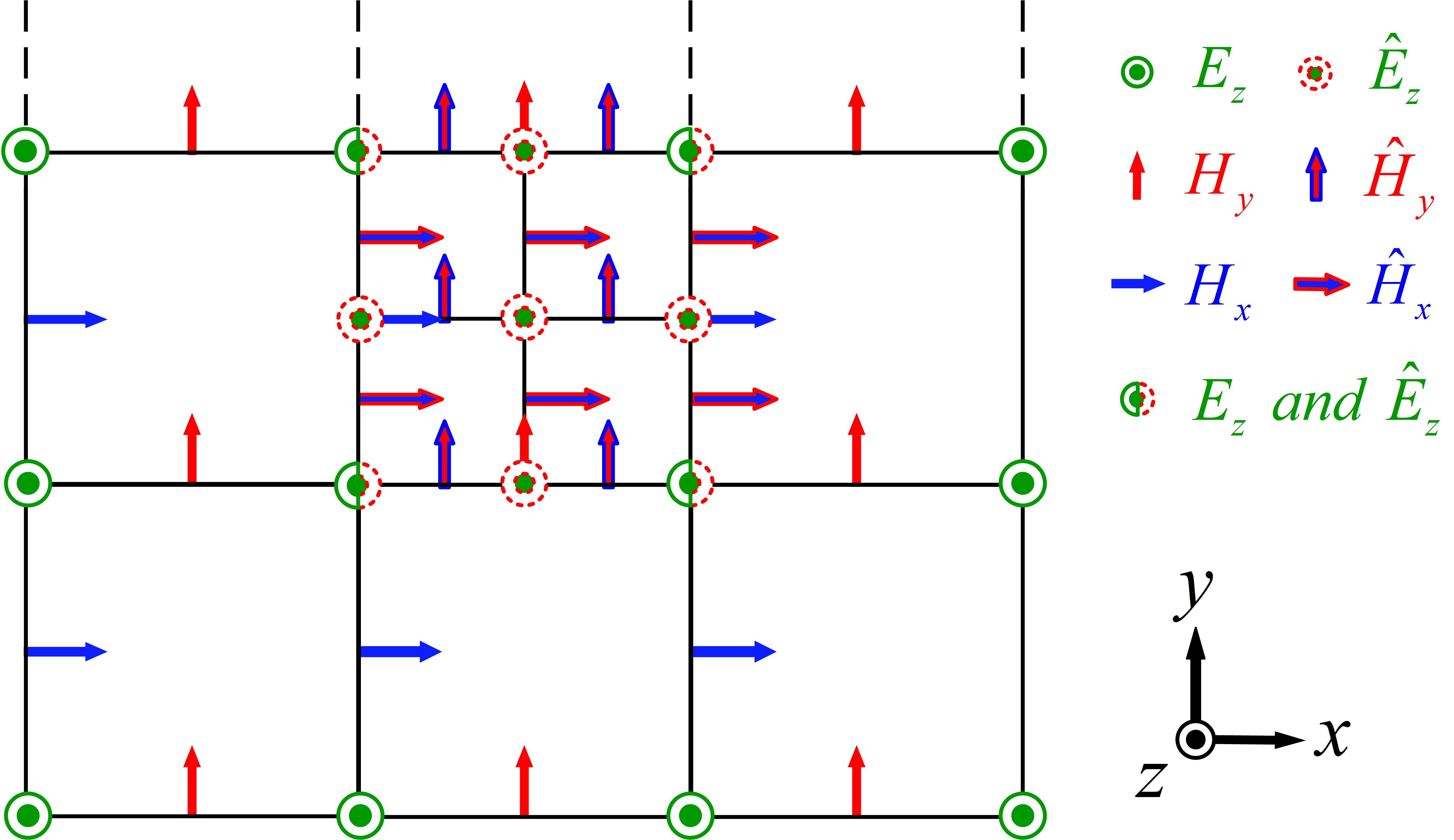}}	
	\caption{The Yee’s grids used in the proposed SBP-SAT FDTD method under the two-dimensional TM assumption and field locations.
	}
\label{figure1}
\end{figure} 

\subsection{The Energy in the Continuous Space}
Without loss of generality, a lossless, linear, isotropic and homogenous medium is considered. Therefore, the Maxwell’s equations under the TM assumption can be expressed as
\begin{subequations} \label{1} 
\begin{align}
&\frac{{\partial {H_x}}}{{\partial t}} = - \frac{1}{\mu} \frac{{\partial {E_z}}}{{\partial y}},\\
&\frac{{\partial {H_y}}}{{\partial t}} = \frac{1}{\mu} \frac{{\partial {E_z}}}{{\partial x}},\\
&\frac{{\partial {E_z}}}{{\partial t}} = \frac{1}{\varepsilon} \left(\frac{{\partial {H_y}}}{{\partial x}} - \frac{{\partial {H_x}}}{{\partial y}} \right),
\end{align}
\end{subequations}
where $\varepsilon$ and $\mu$  are the permittivity and permeability of the medium, respectively. 

The energy of (\ref{1}) in the continuous physical domain \cite{22} can be expressed as
\begin{equation}\label{2} 
	\mathcal{E}=\frac{1}{2} \int_{S}\left(\varepsilon E_{z}{ }^{2}+\mu H_{y}{ }^{2}+\mu H_{x}{ }^{2}\right) d s,
\end{equation}
where $S$ is the overall domain. A rectangular domain used in the FDTD method is considered in this paper. 

We assume that fields are differentiable in the domain. Then, by taking the derivative of (\ref{2}) with respect to time and applying the divergence theorem \cite{24}, we get
\begin{equation}\label{3}
	\frac{d \mathcal{E}}{d t}=\int_{S}\left(\varepsilon E_{z} \frac{\partial E_{z}}{\partial t}+\mu H_{x} \frac{\partial H_{x}}{\partial t}+\mu H_{y} \frac{\partial H_{y}}{\partial t}\right) d s.
\end{equation}

By substituting (\ref{1}) into (\ref{3}), we obtain
\begin{equation}\label{4}
	\begin{aligned}
		\frac{d \mathcal{E}}{d t} &=-\oint_{\partial S} \mathbf{S} \cdot \mathbf{n} d l \\
		&=\int_{\partial S} E_{z} H_{x} d x-\int_{\partial N} E_{z} H_{x} d x \\
		&+\int_{\partial E} E_{z} H_{y} d y-\int_{\partial W} E_{z} H_{y} d y,
	\end{aligned}
\end{equation}
where $\mathbf{S}=\left(E_{z} H_{y},-E_{z} H_{x}, 0\right)$ denotes the Poynting vector in the two-dimensional TM mode, $\mathbf{n}$ is the unit normal vector pointing outwards $S$, $\partial S$ denotes the contour of $S$, and $\partial S$, $\partial N$, $\partial W$ and $\partial E$ are the south, north, west and east boundaries of $S$, respectively. It is easy to find that the derivative of the energy with respect to time, namely the power, is fully determined by the electromagnetic fields on the boundaries in the continuous physical domain, which is determined by the Poynting theorem \cite{25}. In this paper, the SBP operators are carefully designed to mimic (\ref{3}) on the staggered Yee’s grid, and then we use those SBP operators in conjugate with the SAT techniques to solve (\ref{1}), especially in the multiple mesh blocks with different mesh sizes. 

\subsection{The SBP Operator in the One-Dimensional Space}
In this subsection, we construct the SBP discrete system in the two-dimensional transverse magnetic (TM) mode. To well define the SBP operators in the staggered Yee’s grid, the one-dimensional staggered grid is first considered. The grids ${\bf{x}}_-$ and ${\bf{x}}_+$ are defined as
\begin{equation}\label{5}
	\begin{aligned}
		&\mathbf{x}_{-}=\left(x_{0}, x_{1}, \cdots, x_{N}\right), \\
		&\mathbf{x}_{+}=\left(x_{1 / 2}, x_{3 / 2}, \cdots, x_{N-1/2}\right),
	\end{aligned}
\end{equation}
where the subscript of point $x_i$ denotes its coordinate as $ih$ with uniform cell size $h$. $\bf{E}$ nodes are assigned on $\mathbf{x}_{-}$ and $\bf{H}$ nodes are placed on $\bf{x}_{+}$. $\mathbb{D}_+$ and $\mathbb{D}_-$ are the discrete differential matrices with dimensions of $N_{+} \times N_{-}$, $N_{-} \times N_{+}$, respectively, which are used to approximate the $\partial H / \partial x$ and  $\partial E / \partial x$. $\mathbb{P}_+$ and $\mathbb{P}_-$ are the norm matrices with dimensions of $N_{+} \times N_{+}$, $N_{-} \times N_{-}$, respectively, which correspond to the quadrature weights associated with the nodes on $\bf{x}_{-}$ and $\bf{x}_{+}$ \cite{23}. We can further define the following operators
\begin{equation}\label{6}
	\mathbb{Q}_{+}=\mathbb{P}_{+} \mathbb{D}_{+}, \mathbb{Q}_{-}=\mathbb{P}_{-} \mathbb{D}_{-} .
\end{equation}
With the definition of the SBP operator \cite{21}, we have
\begin{equation}\label{7}
	\mathbb{Q}_{+}+\mathbb{Q}^{T}=\mathbb{B},
\end{equation}
where $\mathbb{B}=\operatorname{diag}[-1,0 \ldots 0,1]$.

However, one difficulty to meet the SBP property in (\ref{7}) is that the magnetic fields on the boundaries are not defined in the Yee’s grid. In \cite{1}, we added additional nodes to overcome this issue. In this paper, the extrapolation method, which is defined as the project operator \cite{23}, is used to calculate the magnetic fields on the boundaries. According to \cite{21}, we modified $\mathbb{B}$ as
\begin{equation}\label{8}
	\mathbb{B}=-e_{L}\left(\mathcal{P}_{+}^{L}\right)^{T}+e_{R}\left(\mathcal{P}_{+}^{R}\right)^{T},
\end{equation}
where $e_L$ and $e_R$ are column vectors, which are defined as
\begin{equation}\label{9}
	e_{L}=[1,0, \ldots, 0,0]^{T},
\end{equation}
\begin{equation}\label{10}
	e_{R}=[0,0, \ldots, 0,1]^{T},
\end{equation}
and $\mathcal{P}_{+}^{L}$, $\mathcal{P}_{+}^{R}$ are boundary projection operators \cite{23}. Then, the magnetic fields on the boundaries are $\left(\mathcal{P}_{+}^{L}\right)^{T} \mathbf{H}$ and $\left(\mathcal{P}_{+}^{R}\right)^{T} \mathbf{H}$, respectively, which lead to the well-defined SBP operators to mimic the energy estimate of (\ref{1}) in the staggered Yee’s grid.

In this paper, the second-order central finite-difference scheme are used. To meet the accuracy requirement of the SBP operators, we can define the following matrices
\begin{subequations} \label{11} 
	\begin{align}
&\mathbb{P}_{-}=\operatorname{diag}\left[b_{1}, 1, \ldots, 1, b_{1}\right],\\
&\mathbb{P}_{+}=\operatorname{diag}\left[a_{1}, a_{2}, 1, \ldots, 1, a_{2}, a_{1}\right],\\
&	\mathbb{Q}_{+}=\left[\begin{array}{cccccccc}
			\alpha_{11} & \alpha_{12} & 0 & & & & & \\
			\alpha_{21} & 1 & 0 & & & & & \\
			& -1 & 1 & & & & & \\
			& & -1 & 1 & & & & \\
			& & & \ddots & \ddots & & & \\
			& & & & -1 & 1 & & \\
			& & & & & -1 & 1 & \\
			& & & & & 0 & -1 & -\alpha_{21} \\
			& & & & & 0 & -\alpha_{12} & -\alpha_{11}
		\end{array}\right],\\
&\mathbb{Q}_{-}=\left[\begin{array}{cccccccc}
			\beta_{11} & -\alpha_{21} & 0 & & & & & \\
			& -1 & 1 & & & & & \\
			& & -1 & 1 & & & & \\
			& & & \ddots & \ddots & & & \\
			& & & & -1 & 1 & & \\
			& & & & & -1 & 1 & \\
			& & & & & 0 & \alpha_{21} & -\beta_{11}
		\end{array}\right].
	\end{align}
\end{subequations}

It should be noted that the projection operators $\mathcal{P}_{+}^{L}$ and $\mathcal{P}_{+}^{R}$ are the first-order accuracy, which keep its accuracy consistent with that of the differential operator near the boundaries. To calculate all the parameters defined in (\ref{11}a)-(\ref{11}d), the accuracy relationship should be satisfied
\begin{subequations} \label{12} 
	\begin{align}
&\mathbb{D}_- \mathbf{x}_{+}^{k}=k \mathbf{x}_{-}^{k-1},\\
&\mathbb{D}_+ \mathbf{x}_{-}^{k}=k \mathbf{x}_{+}^{k-1},
	\end{align}
\end{subequations}
where $k = 0, 1$.

By substituting (\ref{11}a)-(\ref{11}d) into (\ref{12}a) and (\ref{12}b), we can obtain several matrix equations and can finally obtain $a_1=1/2$, $a_2=1$, $b_1=1$, $\alpha_{11}=-1$, $\alpha_{21}=1$, $\alpha_{12}=-1$, $\beta_{11}=-1$. With those carefully defined discrete operators, the SBP-SAT FDTD method to solve (\ref{1}) can be derived.

%

\subsection{The SBP-SAT FDTD Method for the Maxwell’s Equations}  
We consider the Yee’s grid with the uniform cell size $h$ in the two-dimensional space in Fig. \ref{figure1}. $N_x$ and $N_y$ cells are included in the $x$ and $y$ direction, respectively. It is easy to find that the two-dimensional grids for the electric and magnetic fields can be constructed through the tensor product of its corresponding one-dimensional grids. As shown in Fig. \ref{figure1}, all the corresponding two-dimensional operators can be obtained through the Kronecker product of the one-dimensional operator and the identity operators with appropriate dimensions. The semi-discrete forms of (\ref{1}) can be expressed as
\begin{subequations} \label{14} 
	\begin{align}
	&\frac{d \mathbf{E}_{z}}{d t}=\frac{1}{\varepsilon}\left(\mathbb{D}_{x-} \otimes \mathbb{I}_{y}\right) \mathbf{H}_{y}-\frac{1}{\varepsilon}\left(\mathbb{I}_{x} \otimes \mathbb{D}_{y-}\right) \mathbf{H}_{x},\\
	&\frac{d \mathbf{H}_{y}}{d t}=\frac{1}{\mu}\left(\mathbb{D}_{x+} \otimes \mathbb{I}_{y}\right) \mathbf{E}_{z},\\
	&\frac{d \mathbf{H}_{x}}{d t}=-\frac{1}{\mu}\left(\mathbb{I}_{x} \otimes \mathbb{D}_{y-}\right) \mathbf{E}_{z},
	\end{align}
\end{subequations}		
where $\mathbf{H}_{x}$ and $\mathbf{H}_{y}$ are the column vectors with dimensions of $N_x N_y \times 1$, $N_x N_y \times 1$, respectively, which collect all the magnetic field nodes in the $x$ and $y$ direction, respectively, in a column-wise fashion. $\mathbf{E}_{z}$ is a column vector for all the electric field nodes in a column-wise fashion. $\mathbb{I}_{x}$, $\mathbb{I}_{y}$ are the identity matrices with dimensions of $N_x \times N_x$, $N_y \times N_y$, respectively. $\mathbb{D}_{x+}$, $\mathbb{D}_{x-}$, $\mathbb{D}_{y+}$ and $\mathbb{D}_{y-}$ are the norm matrices with dimensions of $N_{x+} \times N_{x-}$, $N_{x-} \times N_{x+}$, $N_{y+} \times N_{y-}$ and $N_{y-} \times N_{y+}$, respectively.

The perfect electrically conducting (PEC) boundary conditions, $E_{z}=0$, are assumed on the four outer boundaries. In the SBP-SAT FDTD method, the PEC boundary condition is weakly enforced through the SAT techniques \cite{26}. (\ref{14}) is expressed as
\begin{subequations} \label{15} 
	\begin{align}
		&\frac{d \mathbf{E}_{z}}{d t}=\frac{1}{\varepsilon}\left(\mathbb{D}_{x-} \otimes \mathbb{I}_{y}\right) \mathbf{H}_{y}-\frac{1}{\varepsilon}\left(\mathbb{I}_{x} \otimes \mathbb{D}_{y-}\right) \mathbf{H}_{x}, \\
		&\frac{d \mathbf{H}_{y}}{d t}=\frac{1}{\mu}\left(\mathbb{D}_{x+} \otimes \mathbb{I}_{y}\right) \mathbf{E}_{z} \notag \\
		&\quad+\sigma_{E}\left(\mathbb{P}_{x+} \otimes \mathbb{P}_{y-}\right)^{-1}\left(\mathcal{P}_{+}^{x_{R}} \otimes \mathbb{I}_{y}\right) \mathbb{P}_{y-}\left(\left(e_{x_{R}}\right)^{T} \otimes \mathbb{I}_{y}\right) \mathbf{E}_{z} \notag \\
		&\quad+\sigma_{W}\left(\mathbb{P}_{x+} \otimes \mathbb{P}_{y-}\right)^{-1}\left(\mathcal{P}_{+}^{x_{L}} \otimes \mathbb{I}_{y}\right) \mathbb{P}_{y-}\left(\left(e_{x_{L}}\right)^{T} \otimes \mathbb{I}_{y}\right) \mathbf{E}_{z}, \\
		&\frac{d \mathbf{H}_{x}}{d t}=-\frac{1}{\mu}\left(\mathbb{I}_{x} \otimes \mathbb{D}_{y+}\right) \mathbf{E}_{z} \notag \\
		&\quad+\sigma_{S}\left(\mathbb{P}_{x-} \otimes \mathbb{P}_{y+}\right)^{-1}\left(\mathbb{I}_{x} \otimes \mathcal{P}_{+}^{y_{L}}\right) \mathbb{P}_{x-}\left(\mathbb{I}_{x} \otimes\left(e_{y_{L}}\right)^{T}\right) \mathbf{E}_{z} \notag \\
		&\quad+\sigma_{N}\left(\mathbb{P}_{x-} \otimes \mathbb{P}_{y+}\right)^{-1}\left(\mathbb{I}_{x} \otimes \mathcal{P}_{+}^{y_{R}}\right) \mathbb{P}_{x-}\left(\mathbb{I}_{x} \otimes\left(e_{y_{R}}\right)^{T}\right) \mathbf{E}_{z},
	\end{align}
\end{subequations}
where $\sigma_{W}$, $\sigma_{E}$, $\sigma_{S}$ and $\sigma_{N}$ are free parameters to ensure the stability of the proposed SBP-SAT FDTD method.

The energy of (\ref{15}) is defined as
\begin{equation} \label{16}
	\begin{aligned}
		\mathcal{E} &=\frac{1}{2} \mathbf{E}_{z}{ }^{T}\left(\mathbb{P}_{x-} \otimes \mathbb{P}_{y-}\right) \mathbf{E}_{z}+\frac{1}{2} \mathbf{H}_{y}{ }^{T}\left(\mathbb{P}_{x+} \otimes \mathbb{P}_{y-}\right) \mathbf{H}_{y} \\
		&+\frac{1}{2} \mathbf{H}_{x}{ }^{T}\left(\mathbb{P}_{x-} \otimes \mathbb{P}_{y+}\right) \mathbf{H}_{x},
	\end{aligned}
\end{equation}
which mimics (\ref{4}) in the discrete space.

By substituting (\ref{15}) into (\ref{16}) and taking the derivative with respect to time, we obtain
\begin{equation}
	\begin{aligned} \label{17}
		\frac{d \mathcal{E}}{d t} &=\left(\sigma_{W}+1\right)\left(\mathbf{H}_{y}\right)^{T}\left(\mathcal{P}_{+}^{x_{L}} \otimes \mathbb{I}_{y}\right) \mathbb{P}_{y-}\left(\left(e_{x_{L}}\right)^{T} \otimes \mathbb{I}_{y}\right) \mathbf{E}_{z} \\
		&+\left(\sigma_{E}-1\right)\left(\mathbf{H}_{y}\right)^{T}\left(\mathcal{P}_{+}^{x_{R}} \otimes \mathbb{I}_{y}\right) \mathbb{P}_{y-}\left(\left(e_{x_{R}}\right)^{T} \otimes \mathbb{I}_{y}\right) \mathbf{E}_{z} \\
		&+\left(\sigma_{S}+1\right)\left(\mathbf{H}_{x}\right)^{T}\left(\mathbb{I}_{x} \otimes \mathcal{P}_{+}^{y_{L}}\right) \mathbb{P}_{x-}\left(\mathbb{I}_{x} \otimes\left(e_{y_{L}}\right)^{T}\right) \mathbf{E}_{z} \\
		&+\left(\sigma_{N}-1\right)\left(\mathbf{H}_{x}\right)^{T}\left(\mathbb{I}_{x} \otimes \mathcal{P}_{+}^{y_{R}}\right) \mathbb{P}_{x-}\left(\mathbb{I}_{x} \otimes\left(e_{y_{R}}\right)^{T}\right) \mathbf{E}_{z}.
	\end{aligned}
\end{equation}
It can be found that (\ref{17}) includes four terms, which correspond to the counterparts on the right hand side (RHS) of (\ref{4}).

One sufficient condition $d \mathcal{E} / d t=0$ can guarantee that the semi-discrete system of (\ref{17}) is stable. Therefore, one option is that each term on the RHS of (\ref{17}) vanishes. Therefore, we obtain $\sigma_{W}=-1$, $\sigma_{E}=1$, $\sigma_{S}=-1$, $\sigma_{N}=1$. Those parameters are used in the simulations of Section V, otherwise stated.

When the second-order central finite-difference scheme is used in the temporal domain, we have the time-marching formulations for the proposed SBP-SAT FDTD method
\begin{subequations}\label{18}
	\begin{align} 
	\mathbf{E}_{z}^{n+1 / 2}
	&=\mathbf{E}_{z}^{n-1 / 2}+\frac{\Delta t}{\varepsilon_{0}}\left(\mathbb{P}_{x-} \otimes \mathbb{P}_{y-}\right)^{-1} \notag \\
	& \times \left(\left(\mathbb{D}_{x-} \otimes \mathbb{I}_{y}\right) \mathbf{H}_{y}^{n}-\left(\mathbb{I}_{x} \otimes \mathbb{D}_{y-}\right) \mathbf{H}_{x}^{n}\right), \\
	\mathbf{H}_{y}^{n+1}
	&=\mathbf{H}_{y}^{n}+\frac{\Delta t}{\mu_{0}}\left(\mathbb{D}_{x+} \otimes \mathbb{I}_{y}\right) \mathbf{E}_{z}^{n+1 / 2} \notag \\
	&+\frac{\Delta t \cdot \sigma_{E}}{\mu_{0}}\left(\mathbb{P}_{x+} \otimes \mathbb{P}_{y-}\right)^{-1}\left(\mathcal{P}_{+}^{x_{R}} \otimes \mathbb{I}_{y}\right) \notag \\
	&\times \mathbb{P}_{y-} \left(\left(e_{x_{R}}\right)^{T} \otimes \mathbb{I}_{y}\right) \mathbf{E}_{z}^{n+1 / 2} \notag \\
	&+\frac{\Delta t \cdot \sigma_{W}}{\mu_{0}}\left(\mathbb{P}_{x+} \otimes \mathbb{P}_{y-}\right)^{-1}\left(\mathcal{P}_{+}^{x_{L}} \otimes \mathbb{I}_{y}\right) \notag \\
	&\times \mathbb{P}_{y-} \left(\left(e_{x_{L}}\right)^{T} \otimes \mathbb{I}_{y}\right) \mathbf{E}_{z}^{n+1 / 2},\\
	\mathbf{H}_{x}^{n+1} &=\mathbf{H}_{x}^{n}-\frac{1}{\mu}\left(\mathbb{I}_{x} \otimes \mathbb{D}_{y+}\right) \mathbf{E}_{z}^{n+1 / 2} \notag \\
	&+\frac{\Delta t \cdot \sigma_{S}}{\mu}\left(\mathbb{P}_{x-} \otimes \mathbb{P}_{y+}\right)^{-1}\left(\mathbb{I}_{x} \otimes \mathcal{P}_{+}^{y_{L}}\right) \notag \\
	&\times \mathbb{P}_{x-}\left(\mathbb{I}_{x} \otimes\left(e_{y_{L}}\right)^{T}\right) \mathbf{E}_{z}^{n+1 / 2} \notag \\
	&+\frac{\Delta t \cdot \sigma_{N}}{\mu}\left(\mathbb{P}_{x-} \otimes \mathbb{P}_{y+}\right)^{-1}\left(\mathbb{I}_{x} \otimes \mathcal{P}_{+}^{y_{R}}\right) \notag \\
	&\times \mathbb{P}_{x-}\left(\mathbb{I}_{x} \otimes\left(e_{y_{R}}\right)^{T}\right) \mathbf{E}_{z}^{n+1 / 2}.
	\end{align}
\end{subequations}

Since the explicit time-marching formulations are used in (17), time steps in the simulations are also constrained by cell sizes and the boundary conditions, namely the CFL condition. It can be derived in a similar way in \cite{1}.

\subsection{Comparison Among the FDTD Method, the SBP-SAT FDTD Method in \cite{1} and the Proposed SBP-FDTD Method} 
In this subsection, we compare the grids used in the traditional FDTD Yee’s grid, our previous work \cite{1} and the proposed SBP-SAT FDTD method.
\begin{figure} 	
	\begin{minipage}[h]{0.48\linewidth}\label{twoa}
		\centerline{\includegraphics[scale=0.9]{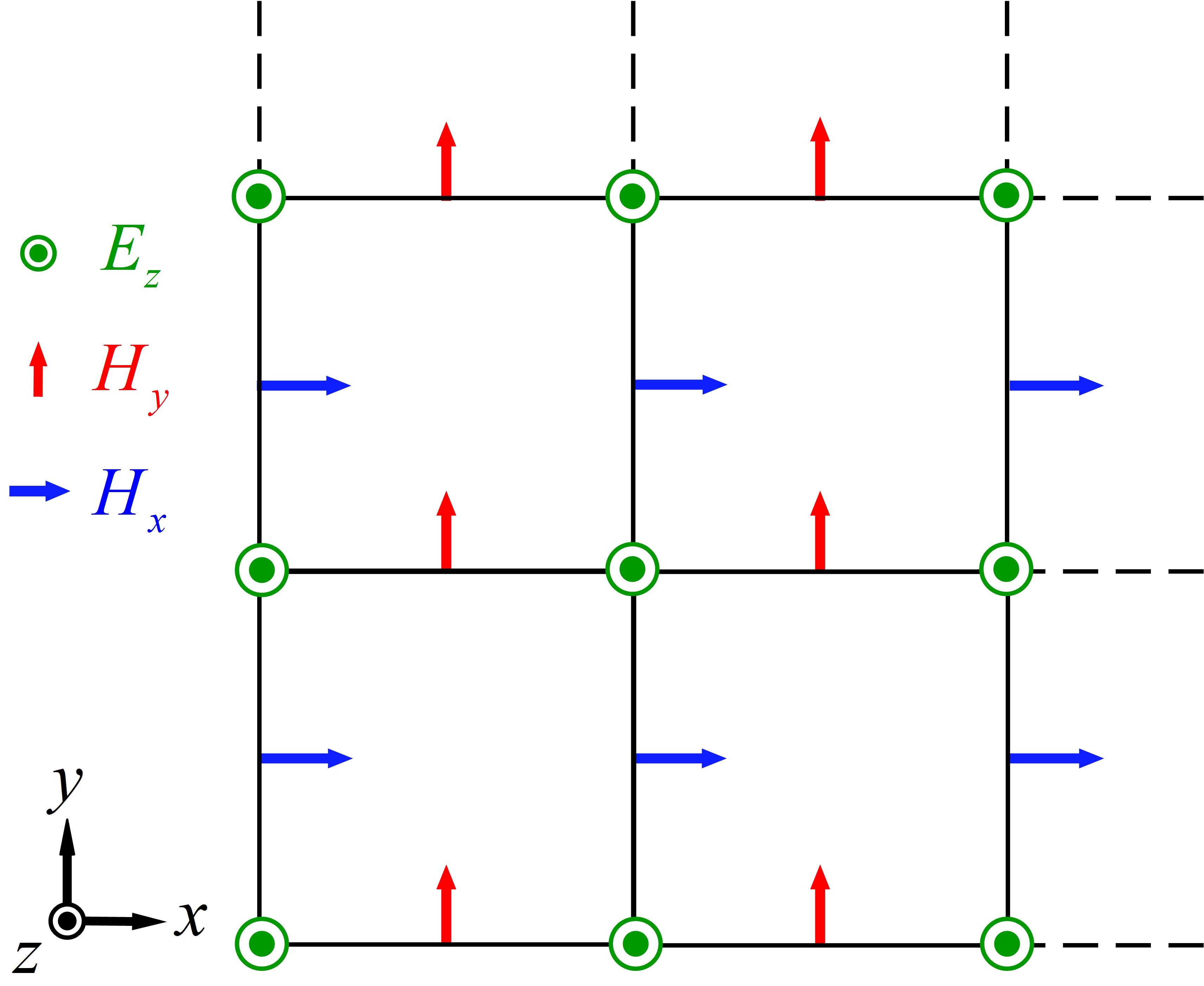}}
		\centerline{(a)}
	\end{minipage}
	\centering
	\begin{minipage}[h]{0.48\linewidth}\label{twob}
		\centerline{\includegraphics[scale=0.9]{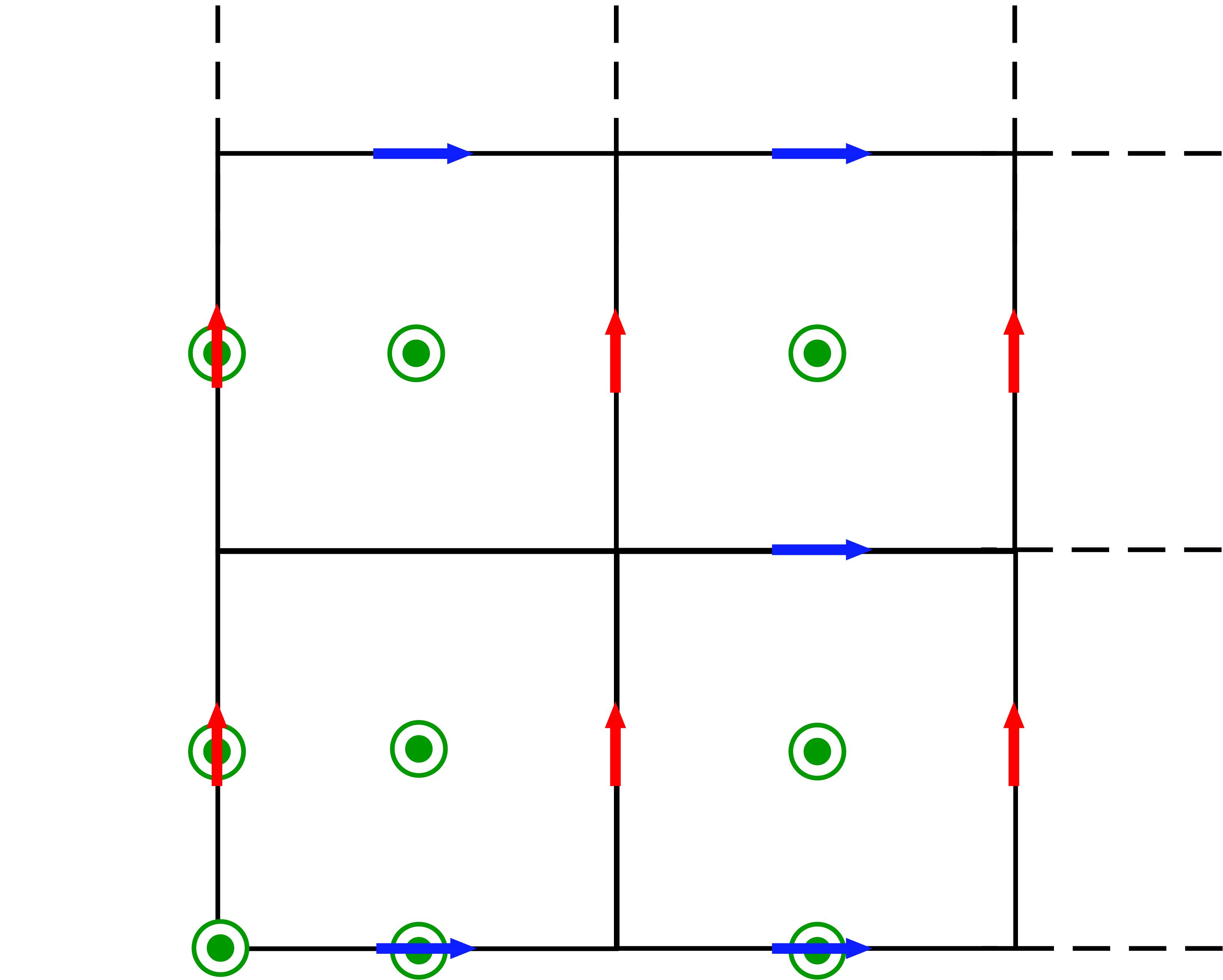}}
		\centerline{(b)}
	\end{minipage}
	\centering
	\caption{Different grids used in the FDTD method and the SBP-SAT FDTD method. (a) Yee’s grids in the FDTD method and the proposed SBP-SAT FDTD method without modifying field components, (b) grids in the SBP-SAT method with modifying field components in \cite{1}.}
	\label{figure2}
\end{figure}

As shown in Fig. \ref{figure2}(a), in the traditional FDTD Yee’s grid and the proposed SBP-SAT FDTD method, magnetic and electric fields are half-cell away from each other, and field components on each grid point depend on the field values at the previous time step and on the four adjacent points. However, the traditional FDTD Yee’s gird does not define the magnetic field on the boundary, so it cannot meet the SBP properties in (\ref{7}).

Fig. \ref{figure2}(b) shows that in the SBP method with modifying field components \cite{1}, several field components are properly added to the boundaries of Yee’s grid to make sure that the discrete operators meet the SBP property. Additional field nodes are added to define magnetic field on the boundaries. 

However, in the proposed SBP-SAT FDTD method, field extrapolation is used to calculate magnetic fields on the boundaries, which satisfies the SBP property on the staggered Yee's grid. By introducing the boundary projection operators \cite{23}, magnetic fields on the boundary can be obtained through the projection of electric fields at the position of half-cell nodes. Therefore, the defined SBP operator can simulate the energy estimation in the traditional Yee’s grids. The grids used in proposed SBP-SAT method is the same as the traditional FDTD Yee’s grid, as shown in Fig. \ref{figure2}(a). 
 
\section{Stable Coupling for Multiple Block Meshes}
\begin{figure}  
	\centerline{\includegraphics[scale=1.1]{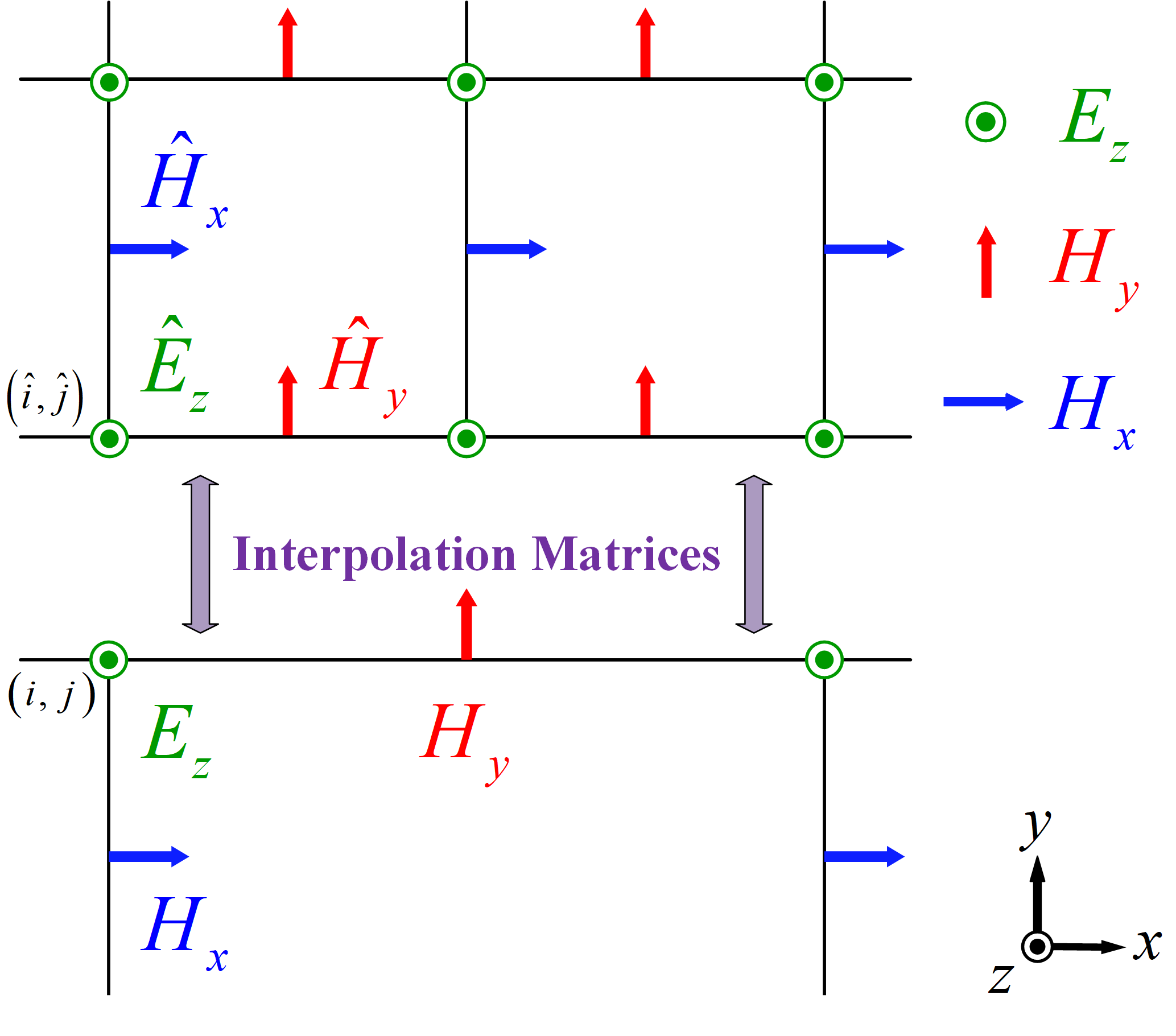}}	
	\caption{Two mesh blocks with cell size ratio 2:1 are horizontally connected.}
	\label{figure3}	
\end{figure}

When the subgridding meshes are used in the practical simulations, the interfaces between mesh blocks with different mesh sizes need to be carefully considered. As shown in Fig. \ref{figure3}, two grids with cell size ratio 2:1 are used to demonstrate the application of the proposed SBP-SAT FDTD method to handle the interface. The two mesh blocks are connected by a horizontal interface. The north boundary of the bottom domain is connected with the south boundary of the top domain. To make the following derivation clear, the quantities with $\widehat{}$ de-note they are defined on the bottom mesh block. Therefore, the boundary conditions on the interface in the continuous domain can be expressed as 
\begin{subequations}\label{19}
	\begin{align} 	
		&\widehat{\mathbf{E}}_{z}=\mathbf{E}_{z}, \\
		&\widehat{\mathbf{H}}_{x}=\mathbf{H}_{x}.
	\end{align}
\end{subequations}

Since $\mathbf{E}_{z}$ nodes are defined on the interfaces, (\ref{19}a) can be directly enforced through the SAT technique. However, $\mathbf{H}_{x}$ components are not defined on the south boundary of the top mesh block and on the north boundary of the bottom mesh block. To address this issue, the project operator will be used to extrapolate $\mathbf{H}_{x}$ on the interface.

Through adding the SAT terms to the semi-discrete formulations in the top domain, fields can be expressed as
\begin{subequations}\label{20} 
	\begin{align}
		\frac{d \widehat{\mathbf{E}}_{z}}{d t}
		&=\frac{1}{\varepsilon}\left(\widehat{\mathbb{D}}_{x-} \otimes \widehat{\mathbb{I}}_{y}\right) \widehat{\mathbf{H}}_{y}-\frac{1}{\varepsilon}\left(\widehat{\mathbb{I}}_{x} \otimes \widehat{\mathbb{D}}_{y-}\right) \widehat{\mathbf{H}}_{x} \notag\\
		&+\widehat{\sigma}_{E_{z}}\left(\widehat{\mathbb{P}}_{x-} \otimes \widehat{\mathbb{P}}_{y-}\right)^{-1}\left(\widehat{\mathbb{I}}_{x} \otimes \widehat{e}_{y_{R}}\right) \widehat{\mathbb{P}}_{x-} \notag\\
		&\times \left[\begin{array}{l}
			\left.\widehat{\mathbb{T}}\left(\left(\mathbb{I}_{x} \otimes\left(\mathcal{P}_{+}^{x_{L}}\right)^{T}\right) \mathbf{H}_{x}\right)\right. \\
			-\left(\widehat{\mathbb{I}}_{x} \otimes\left(\widehat{\mathcal{P}}_{+}^{x_{R}}\right)^{T}\right) \widehat{\mathbf{H}}_{x}
		\end{array}\right],\\
	\frac{d \widehat{\mathbf{H}}_{y}}{d t}
	&=\frac{1}{\mu}\left(\widehat{\mathbb{D}}_{x+} \otimes \widehat{\mathbb{I}}_{y}\right) \widehat{\mathbf{E}}_{z},\\
		\frac{d \widehat{\mathbf{H}}_{x}}{d t}
		&=-\frac{1}{\mu}\left(\widehat{\mathbb{I}}_{x} \otimes \widehat{\mathbb{D}}_{y-}\right) \widehat{\mathbf{E}}_{z}\notag \\
		&+\widehat{\sigma}_{H_{x}}\left(\widehat{\mathbb{P}}_{x-} \otimes \widehat{\mathbb{P}}_{y+}\right)^{-1}\left(\widehat{\mathbb{I}}_{x} \otimes \widehat{\mathcal{P}}_{+}^{x_{R}}\right) \widehat{\mathbb{P}}_{x-}\notag \\
		&\times \left[\begin{array}{l}
			\left.\widehat{\mathbb{T}}\left(\left(\mathbb{I}_{x} \otimes\left(e_{y_{L}}\right)^{T}\right) \mathbf{E}_{z}\right) \right.\\
			-\left(\widehat{\mathbb{I}}_{x} \otimes\left(\widehat{e}_{y_{R}}\right)^{T}\right) \widehat{\mathbf{E}}_{z}
		\end{array}\right],
\end{align}
\end{subequations}
and in the bottom domain, we have
\begin{subequations}\label{21} 
	\begin{align}
		\frac{d \mathbf{E}_{z}}{d t}
		&=\frac{1}{\varepsilon}\left(\mathbb{D}_{x-} \otimes \mathbb{I}_{y}\right) \mathbf{H}_{y}-\frac{1}{\varepsilon}\left(\mathbb{I}_{x} \otimes \mathbb{D}_{y-}\right) \mathbf{H}_{x} \notag \\
		&+\sigma_{E_{z}}\left(\mathbb{P}_{x-} \otimes \mathbb{P}_{y-}\right)^{-1}\left(\mathbb{I}_{x} \otimes e_{y_{L}}\right) \mathbb{P}_{x-}\notag \\
		&\times \left[\begin{array}{l}
			\left(\mathbb{I}_{x} \otimes\left(\mathcal{P}_{+}^{x_{L}}\right)^{T}\right) \mathbf{H}_{x} \\
			\left.-\mathbb{T}\left(\left(\widehat{\mathbb{I}}_{x} \otimes\left(\widehat{\mathcal{P}}_{+}^{x_{R}}\right)^{T}\right) \widehat{\mathbf{H}}_{x}\right)\right.
		\end{array}\right], \\
	\frac{d \mathbf{H}_{y}}{d t}
	&=\frac{1}{\mu}\left(\mathbb{D}_{x+} \otimes \mathbb{I}_{y}\right) \mathbf{E}_{z}, \\
		\frac{d \mathbf{H}_{x}}{d t}
		&=-\frac{1}{\mu}\left(\mathbb{I}_{x} \otimes \mathbb{D}_{y-}\right) \mathbf{E}_{z}\notag  \\
		&+\sigma_{H_{x}}\left(\mathbb{P}_{x-} \otimes \mathbb{P}_{y+}\right)^{-1}\left(\mathbb{I}_{x} \otimes \mathcal{P}_{+}^{x_{L}}\right) \mathbb{P}_{x-}\notag \\
		&\times \left[\begin{array}{l}
			\left(\mathbb{I}_{x} \otimes\left(e_{y_{L}}\right)^{T}\right) \mathbf{E}_{z} \\
			\left.-\mathbb{T}\left(\left(\widehat{\mathbb{I}}_{x} \otimes\left(\widehat{e}_{y_{R}}\right)^{T}\right) \widehat{\mathbf{E}}_{z}\right)\right.
		\end{array}\right],
\end{align}
\end{subequations}
where $\sigma_{E_{z}}$, $\widehat\sigma_{E_{z}}$, $\sigma_{H_{x}}$ and $\widehat\sigma_{H_{x}}$ are free parameters to guarantee the stability of the proposed SBP-SAT FDTD subgridding method. $\widehat{\mathbb{T}}$ is used to interpolate $\mathbf{H}_{x}$ from the top to bottom domain on the interface, and ${\mathbb{T}}$ from the bottom to top domain. In order to simplify the following discussion, the one-dimensional boundary field column vectors are defined as
\begin{subequations}\label{22} 
	\begin{align}
&\widehat{\mathbf{E}}_{x_{l}}=\left(\widehat{\mathbb{I}}_{x} \otimes\left(\widehat{e}_{y_{R}}\right)^{T}\right) \widehat{\mathbf{E}}_{z}, \\
&\mathbf{E}_{x_{l}}=\left(\mathbb{I}_{x} \otimes\left(e_{y_{L}}\right)^{T}\right) \mathbf{E}_{z}, \\
&\widehat{\mathbf{H}}_{x_{l}}=\left(\widehat{\mathbb{I}}_{x} \otimes\left(\widehat{\mathcal{P}}_{+}^{x_{R}}\right)^{T}\right) \widehat{\mathbf{H}}_{x}, \\
&\mathbf{H}_{x_{l}}=\left(\mathbb{I}_{x} \otimes\left(\mathcal{P}_{+}^{x_{L}}\right)^{T}\right) \mathbf{H}_{x}.
\end{align}
\end{subequations}
Therefore, the energy equals the superposition of the energy in the top and bottom domain, which can be expressed as 
\begin{equation}\label{24} 
	\begin{aligned}
		\mathcal{E} &=\frac{1}{2} \mathbf{E}_{z}{ }^{T}\left(\mathbb{P}_{x-} \otimes \mathbb{P}_{y-}\right) \mathbf{E}_{z}+\frac{1}{2} \mathbf{H}_{y}{ }^{T}\left(\mathbb{P}_{x+} \otimes \mathbb{P}_{y-}\right) \mathbf{H}_{y} \\
		&+\frac{1}{2} \mathbf{H}_{x}{ }^{T}\left(\mathbb{P}_{x-} \otimes \mathbb{P}_{y+}\right) \mathbf{H}_{x}+\frac{1}{2} \widehat{\mathbf{E}}_{z}{ }^{T}\left(\widehat{\mathbb{P}}_{x-} \otimes \widehat{\mathbb{P}}_{y-}\right) \widehat{\mathbf{E}}_{z} \\
		&+\frac{1}{2} \widehat{\mathbf{H}}_{y}{ }^{T}\left(\widehat{\mathbb{P}}_{x+} \otimes \widehat{\mathbb{P}}_{y-}\right) \widehat{\mathbf{H}}_{y}+\frac{1}{2} \widehat{\mathbf{H}}_{x}{ }^{T}\left(\widehat{\mathbb{P}}_{x-} \otimes \widehat{\mathbb{P}}_{y+}\right) \widehat{\mathbf{H}}_{x}.
	\end{aligned}
\end{equation}

By taking the derivative of (\ref{24}) with respect to time, we obtain
\begin{equation}
	\begin{aligned}\label{25} 
		\frac{d \mathcal{E}}{d t} &=-\left(1+\widehat{\sigma}_{E_{z}}+\widehat{\sigma}_{H_{x}}\right) \widehat{\mathbf{E}}_{x_{l}} \widehat{\mathbb{P}}_{x-} \widehat{\mathbf{H}}_{x_{l}} \\
		&+\left(1+\sigma_{E_{z}}+\sigma_{H_{x}}\right) \mathbf{E}_{x_{l}} \mathbb{P}_{x-} \mathbf{H}_{x_{l}} \\
		&-\widehat{\mathbf{E}}_{x_{l}}\left(\sigma_{H_{x}}(\mathbb{T})^{T} \mathbb{P}_{x-}-\widehat{\sigma}_{E_{z}} \widehat{\mathbb{P}}_{x-} \widehat{\mathbb{T}}\right) \mathbf{H}_{x_{l}} \\
		&-\mathbf{E}_{x_{l}}\left(\sigma_{E_{z}} \mathbb{P}_{x-} \mathbb{T}-\widehat{\sigma}_{H_{x}}(\widehat{\mathbb{T}})^{T} \widehat{\mathbb{P}}_{x-}\right) \widehat{\mathbf{H}}_{x_{l}}.
	\end{aligned}
\end{equation}

\newcounter{TempEqCnt} 
\setcounter{TempEqCnt}{\value{equation}} 
\setcounter{equation}{25} 
\begin{figure*}[hb] 
	\hrulefill  
	\begin{align}\label{26}
		\widehat{\mathbb{T}}=\left[\begin{array}{lllllllllllll}
			a_{1} & a_{2} & a_{3} & a_{4} & a_{5} & & & & & & & & \\
			& & c_{1} & c_{2} & c_{3} & c_{4} & c_{5} & & & & & & \\
			& & & & \ddots & \ddots & \ddots & \ddots & \ddots & & & & \\
			& & & & & & c_{1} & c_{2} & c_{3} & c_{4} & c_{5} & & \\
			& & & & & & & & a_{1} & a_{2} & a_{3} & a_{4} & a_{5}
		\end{array}\right] 
	\end{align}
\end{figure*}
\setcounter{equation}{\value{TempEqCnt}}

To ensure the long-time stability of the proposed SBP-SAT FDTD method, $d \mathcal{E} / d t=0$ should be satisfied, which implies that no energy dissipation occurs in the computational domain. One option is to make each term vanish. Therefore, we have 
\begin{subequations}\label{26} 
	\begin{align}
		&1+\widehat{\sigma}_{E_{z}}+\widehat{\sigma}_{H_{x}}=0, \\
		&1+\sigma_{E_{z}}+\sigma_{H_{x}}=0, \\
		&\sigma_{H_{x}}-\widehat{\sigma}_{E_{z}}=0, \\
		&\sigma_{E_{z}}-\widehat{\sigma}_{H_{x}}=0.
	\end{align}
\end{subequations}
Therefore, we obtain $\widehat{\sigma}_{E_{z}}=\sigma_{E_{z}}=\sigma_{H_{x}}=\widehat{\sigma}_{H_{x}}=-1 / 2$. Those parameters are used in the simulations, otherwise stated. When the two remaining terms vanish, we impose additional condition as
\begin{equation}\label{27} 
		(\mathbb{T})^{T} \mathbb{P}_{x-}=\widehat{\mathbb{P}}_{x-} \widehat{\mathbb{T}}.
\end{equation}
This is the norm compatible condition, which is also defined in \cite{28}. To ensure the stability of the subgridding method, (\ref{27}) should be satisfied. To calculate the detailed interpolation matrices, the optimization method has to be used to obtain the interpolation matrix. Interested readers can be referred to \cite{1} for more details.

In this paper, the following interpolation matrix is used in our simulations. Assume that the fine grids are applied in the top domain, as shown in Fig. \ref{figure3}, which implies that $\widehat{\mathbb{T}}$ is the interpolation matrix from fine to coarse meshes, $\mathbb{T}$ is the interpolation matrix from coarse to fine meshes. $\widehat{\mathbb{T}}$ can be expressed as (26) at the bottom of this page according to \cite{44}, which is used to interpolate $\mathbf{H}_{x}$ from the top to bottom domain on the interface. It should be noted that the second-order accuracy is for inner nodes and the first-order accuracy is for boundary closure nodes.

An optimization procedure can be used to To calculate the interpolation matrices in \cite{1}. Interested readers are referred to \cite{1} for more details. Then, the upper left corner of $\widehat{\mathbb{T}}$ is given by


\begin{equation}
	\begin{aligned}
		&a_{1}=0.5505, a_{2}=0.5, a_{3}=-0.5505, \\
		&a_{4}=a_{5}=0, c_{1}=c_{5}=-0.0252, \\
		&c_{2}=c_{4}=0.25, c_{3}=0.5505 .
	\end{aligned}
\end{equation}

\section{Numerical Examples and Discuss}
In this section, four numerical examples are carried out to validate the stability, accuracy, and efficiency of the proposed SBP-SAT FDTD method. All codes are written in Matlab and run on a workstation with Intel i5-1135G7 2.4GHz CPU, 16G memory, and they are all run with a single thread for fair comparison.

\subsection{A Metallic Cavity}   
The first example is a rectangular vacuum filled cavity with PEC boundaries. Its resonant frequencies are calculated by the FDTD method, the SBP-SAT FDTD method, and the SBP-SAT FDTD method with subgridding meshes, respectively.

\begin{figure}  
	\centerline{\includegraphics[scale=0.7]{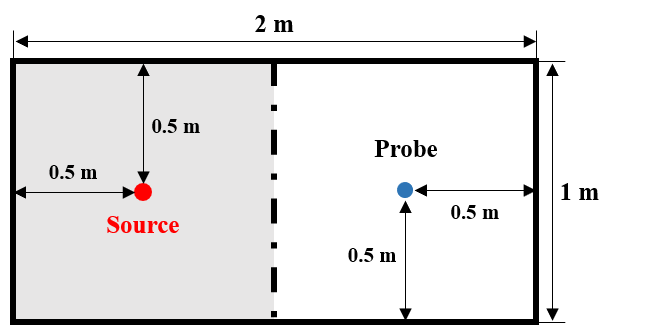}}	
	\caption{Geometrical configuration of the cavity used in the proposed subgridding method, and the source and probe locations are placed at (0.5,0.5) [m], and (1.5,0.5) [m], respectively.}
	\label{figure4}	
\end{figure}

The size of the cavity is 2 m$\times$1 m, as shown in Fig. \ref{figure4}. A Gaussian pulse, located at (0.5,0.5) [m], with the bandwidth of 0.9 GHz is used as the excitation source. A point probe is placed at (1.5,0.5) [m] to record the transient electric fields in the simulations.

Two scenarios are considered in this example. One is that the cavity is discretized with the uniform meshes, with the mesh size $\Delta=2.5 \times 10^{-2}$ m and $\Delta=5 \times 10^{-2}$ m in both the $x$ and $y$ direction. The other one is that the cavity is divided into two domains, as shown in Fig. \ref{figure4}. The left domain is discretized with uniform meshes with mesh size $\Delta=2.5 \times 10^{-2}$ m, and the uniform meshes with the mesh size $\Delta=5 \times 10^{-2}$ m are used in the right domain. The time steps used in the simulations are 29.483 ps, which is 0.99 times of the maximum time steps constrained by the CFL condition in the FDTD method with fine meshes.

\begin{figure} 	
	\begin{minipage}[h]{0.48\linewidth}\label{figure51a}
		\centerline{\includegraphics[scale=0.035]{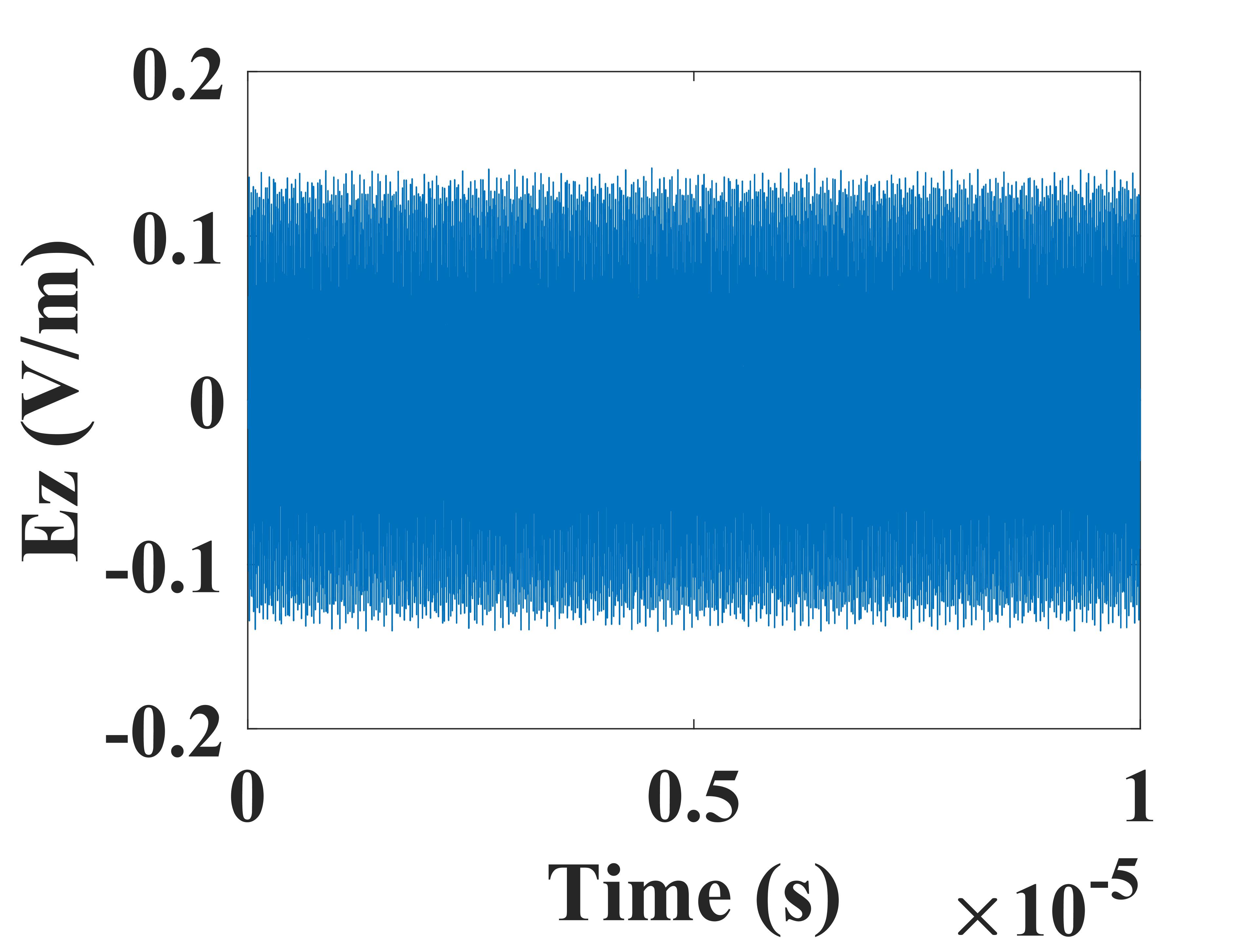}}
		\centerline{(a)}
	\end{minipage}
	\centering
	\begin{minipage}[h]{0.48\linewidth}\label{figure51b}
		\centerline{\includegraphics[scale=0.035]{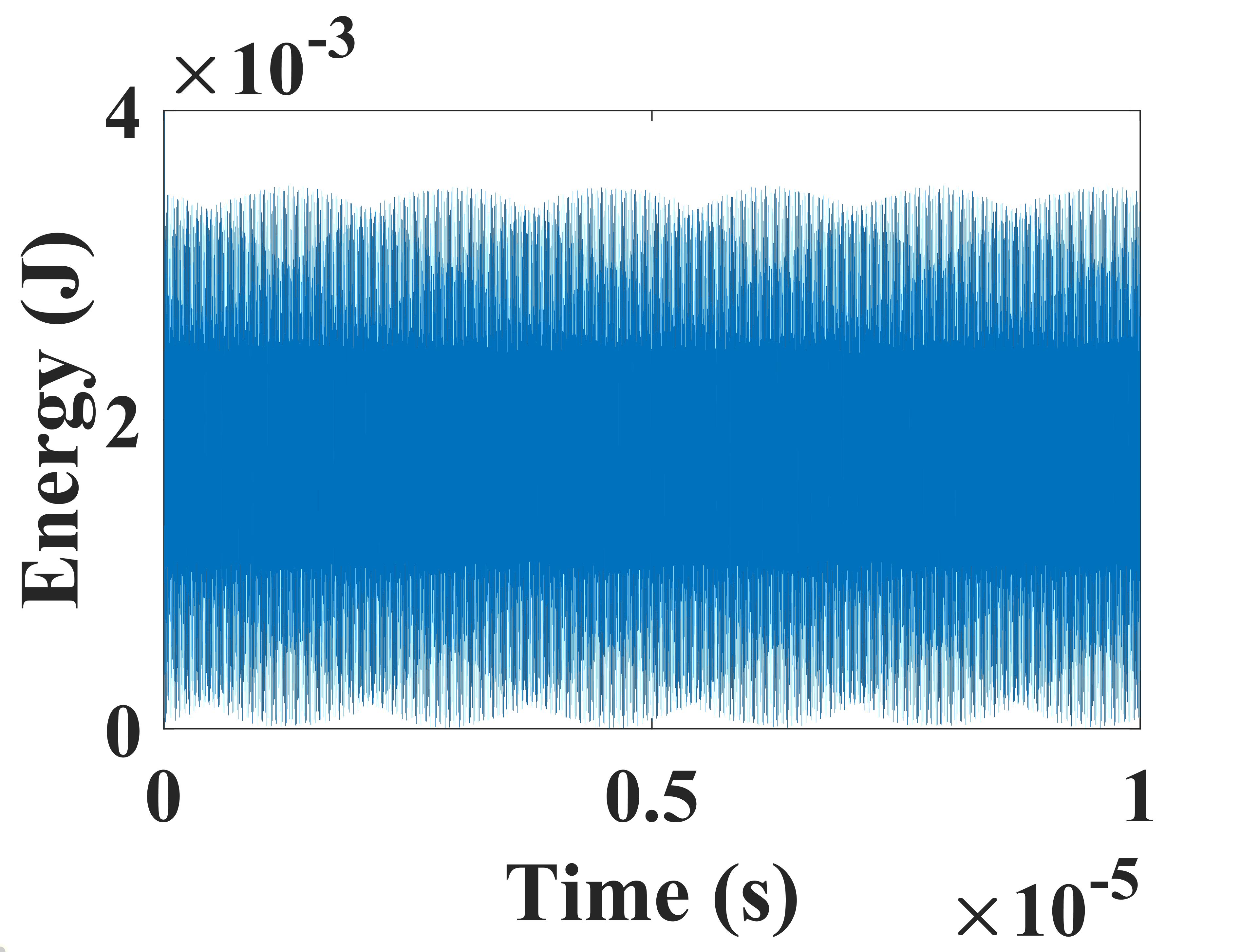}}
		\centerline{(b)}
	\end{minipage}
	\centering
	\caption{(a) The electric fields and (b) the energy verse the time at the probe obtained from the SBP-SAT FDTD method. }
	\label{figure51}
\end{figure}

To investigate the stability of the proposed SBP-SAT FDTD method, we calculate the electric field and the electromagnetic energy in the whole computational domain in Fig. \ref{figure4}. The simulation time is set as $1.0 \times 10^{-5}$ s which results in the total number of time steps is $3.3 \times 10^{7}$. Fig. \ref{figure51} shows the electric field and energy recorded at the probe position. The results show that after a long-time simulation, the electric field and energy obtained by the SBP-SAT FDTD method are stable. Since the discrete partial differential operator used in the proposed SBP-SAT FDTD method satisfies the SBP property, and its long-term stability is theoretically guaranteed.

\begin{figure} 	
	\begin{minipage}[h]{0.48\linewidth}\label{figure52a}
		\centerline{\includegraphics[scale=0.035]{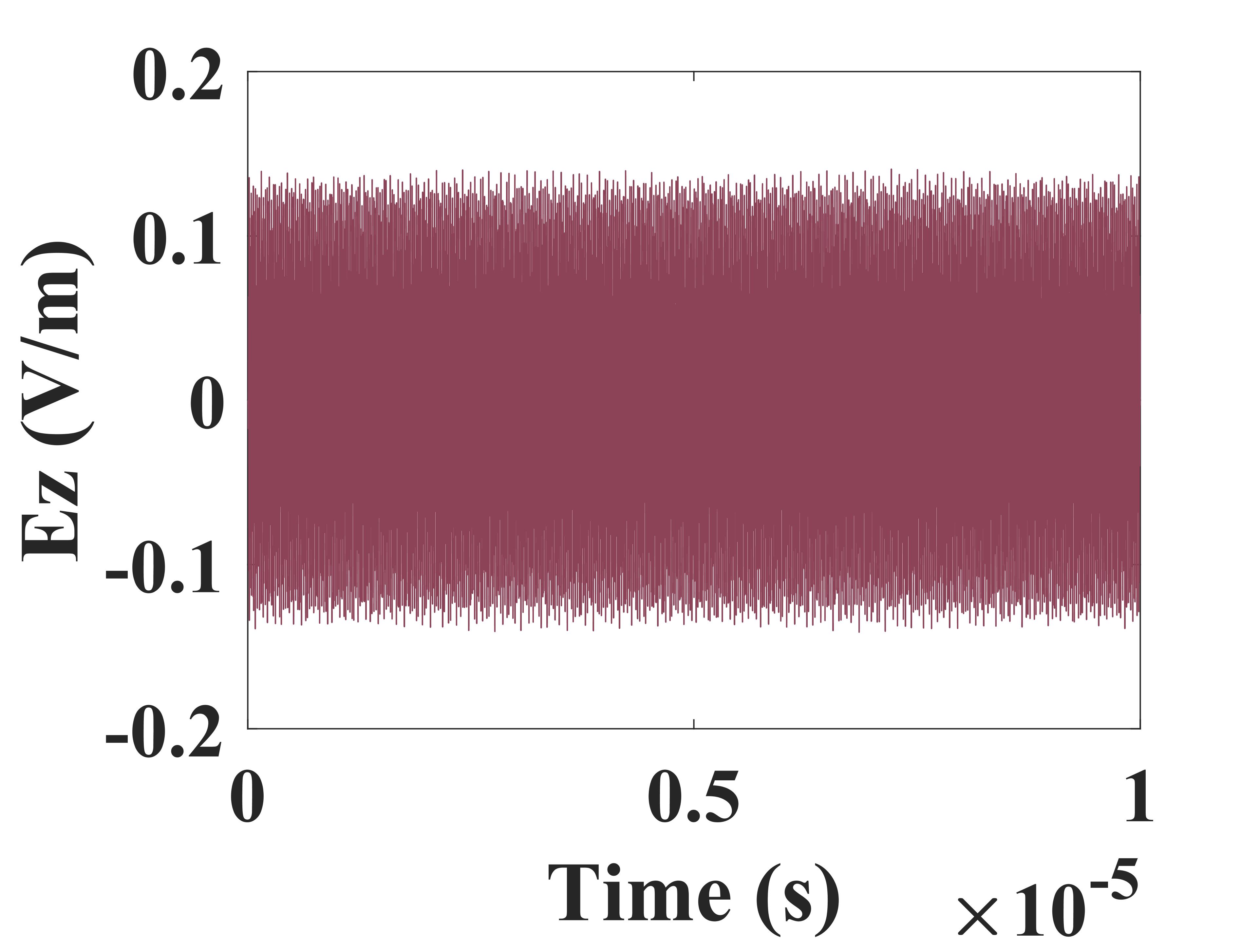}}
		\centerline{(a)}
	\end{minipage}
	\centering
	\begin{minipage}[h]{0.48\linewidth}\label{figure52b}
		\centerline{\includegraphics[scale=0.035]{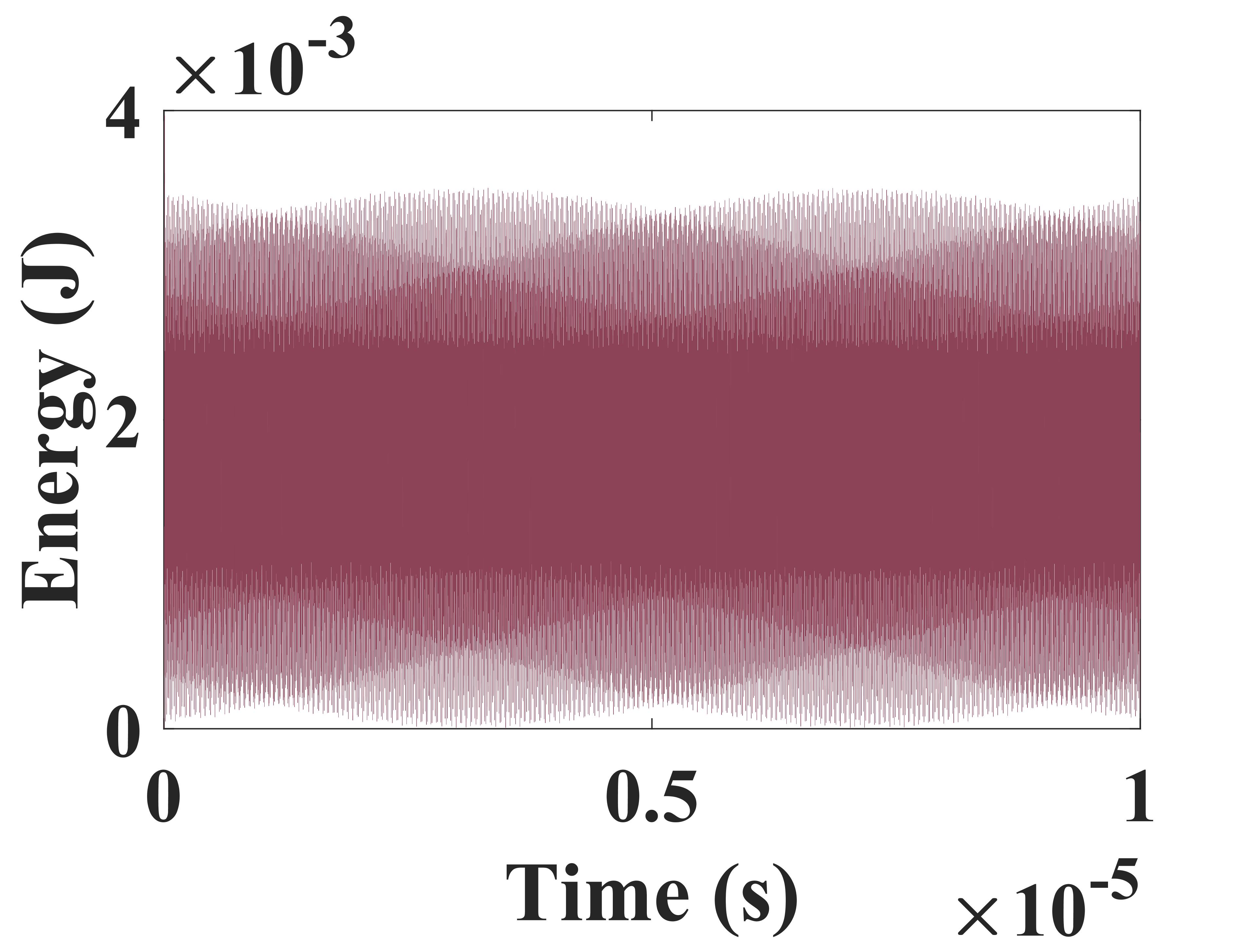}}
		\centerline{(b)}
	\end{minipage}
	\centering
	\caption{(a) The electric fields and (b) the energy verse the time at the probe obtained from the SBP-SAT FDTD subgridding method. }
	\label{figure52}
\end{figure}

Fig. \ref{figure52} shows the electric field and energy obtained from the proposed SBP-SAT FDTD method with the subgridding meshes. The free parameters $\sigma$ and the interpolation matrix in (\ref{25}) are designed to ensure the stability. The electric field and energy are stable after a long-time simulation. Therefore, the stability of our previous analytical analysis is also numerically verified.

\begin{figure}
	\begin{minipage}[h]{0.325\linewidth}\label{fivea}
	\centerline{\includegraphics[scale=0.28]{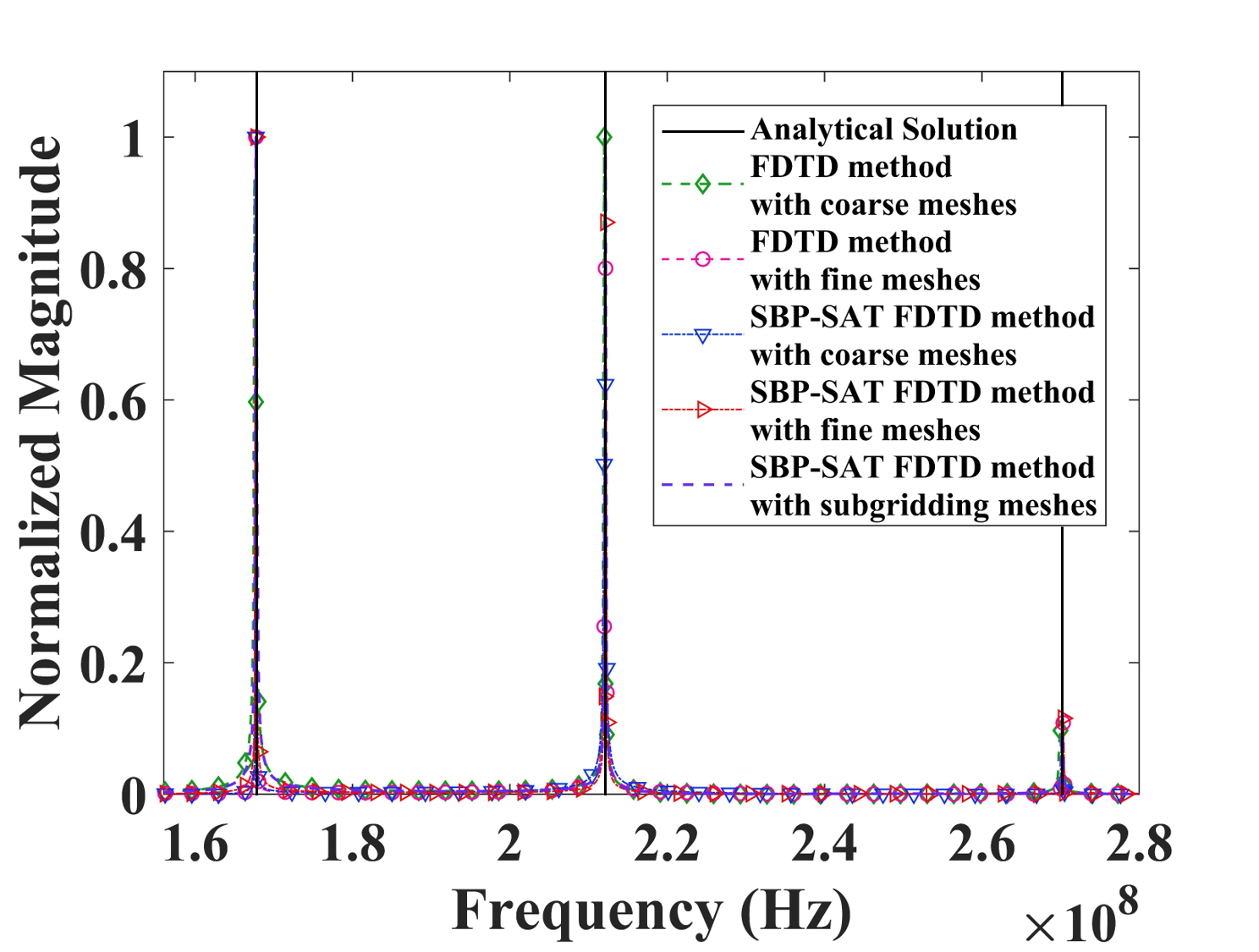}}
	\centerline{(a)}
\end{minipage}
\centering \\
\begin{minipage}[h]{0.325\linewidth}\label{fiveb}
	\centerline{\includegraphics[scale=0.28]{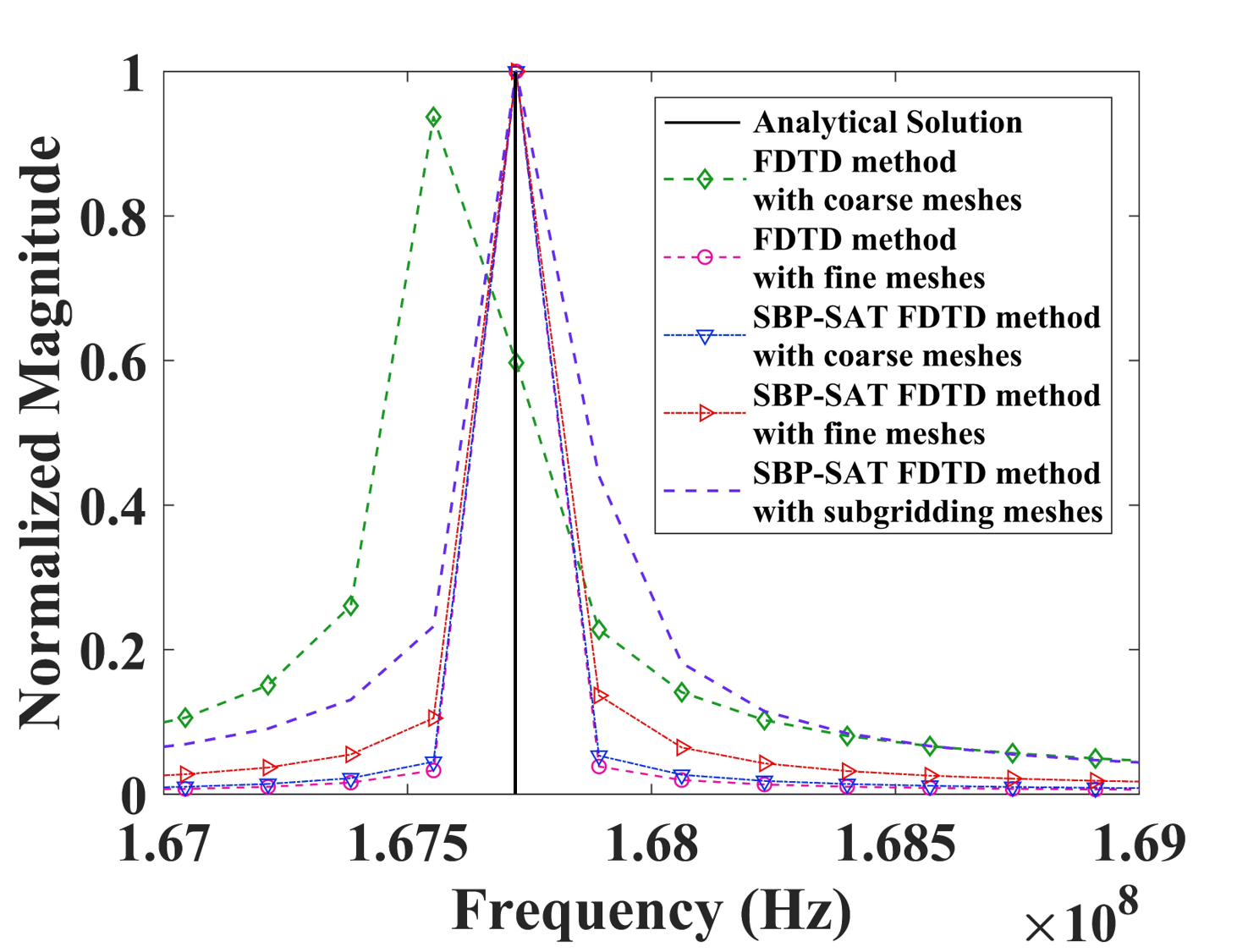}}
	\centerline{(b)}
\end{minipage}
\centering \\
\begin{minipage}[h]{0.325\linewidth}\label{fivec}
	\centerline{\includegraphics[scale=0.28]{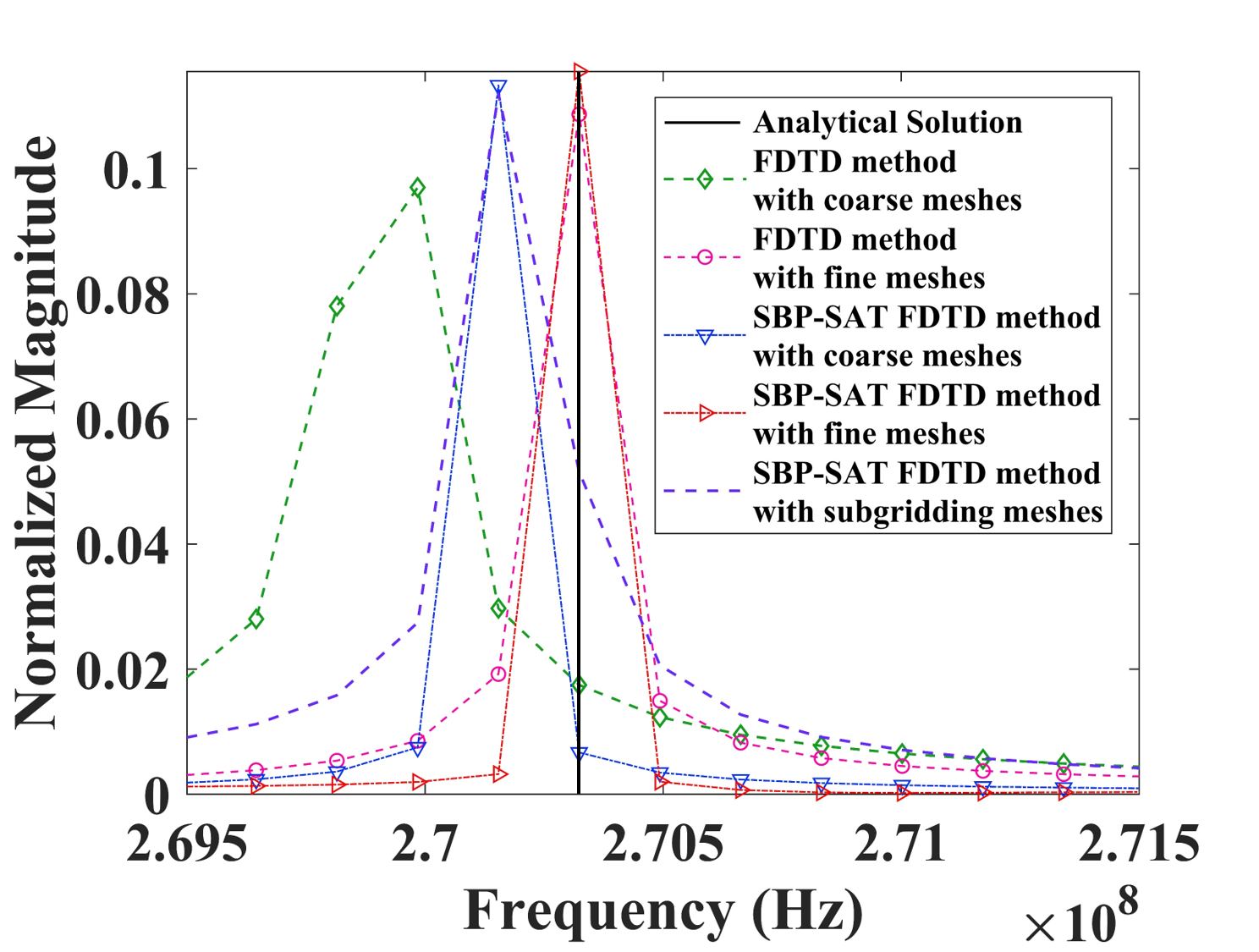}}
	\centerline{(c)}
\end{minipage}
\centering
\caption{The resonant frequencies of TM mode obtained from the FDTD method, the SBP-SAT FDTD method and the proposed SBP-SAT FDTD subgridding method. (a) Three resonant frequencies of the cavity with the frequency range [$1.550 \times 10^{8}$,$2.800 \times 10^{8}$] (Hz), (b) [$1.670 \times 10^{8}$,$1.690 \times 10^{8}$] (Hz), (c) [$2.695 \times 10^{8}$,$2.715 \times 10^{8}$] (Hz), respectively.}
\label{figure5}
\end{figure}

The resonant frequencies of the cavity are calculated by the discrete Fourier transform of the transient electric fields. Results obtained from the FDTD method with fine mesh sizes $\Delta=2 \times 10^{-2}$ m and the analytical one are also plotted as the references as shown in Fig. \ref{figure5}. It is seen that the results obtained by the SBP-SAT FDTD method agree well with the FDTD method with fine grids. In addition, when coarse meshes are used in the computational domain, the results obtained by the two FDTD methods are also similar to each other. Since the proposed SBP-SAT FDTD method and the FDTD method used the second-order difference scheme, the same level of accuracy can be obtained. As shown in Fig. \ref{figure5}, compared with the FDTD method and the SBP-SAT FDTD method with coarse meshes, the accuracy of the SBP-SAT FDTD subgridding method is improved.

\subsection{The SAR Calculation} 
To verify the feasibility and accuracy of the proposed subgridding method, a human head model is considered as a practical test. By calculating the specific absorption rate (SAR) of human head irradiated by the electromagnetic waves, we compare results obtain from the FDTD method with coarse and fine meshes, the SBP-SAT FDTD method with coarse and fine meshes, and the subgridding meshes.

In the $y$ direction, the boundary conditions on the interface in the continuous domain can be expressed as (18). And in the $x$ direction, the conditions can be defined as
\begin{subequations}
	\begin{align} \label{28}	
		&\widehat{\mathbf{E}}_{z}=\mathbf{E}_{z}, \\
		&\widehat{\mathbf{H}}_{y}=\mathbf{H}_{y}.
	\end{align}
\end{subequations}

\begin{figure}
	\centerline{\includegraphics[scale=0.25]{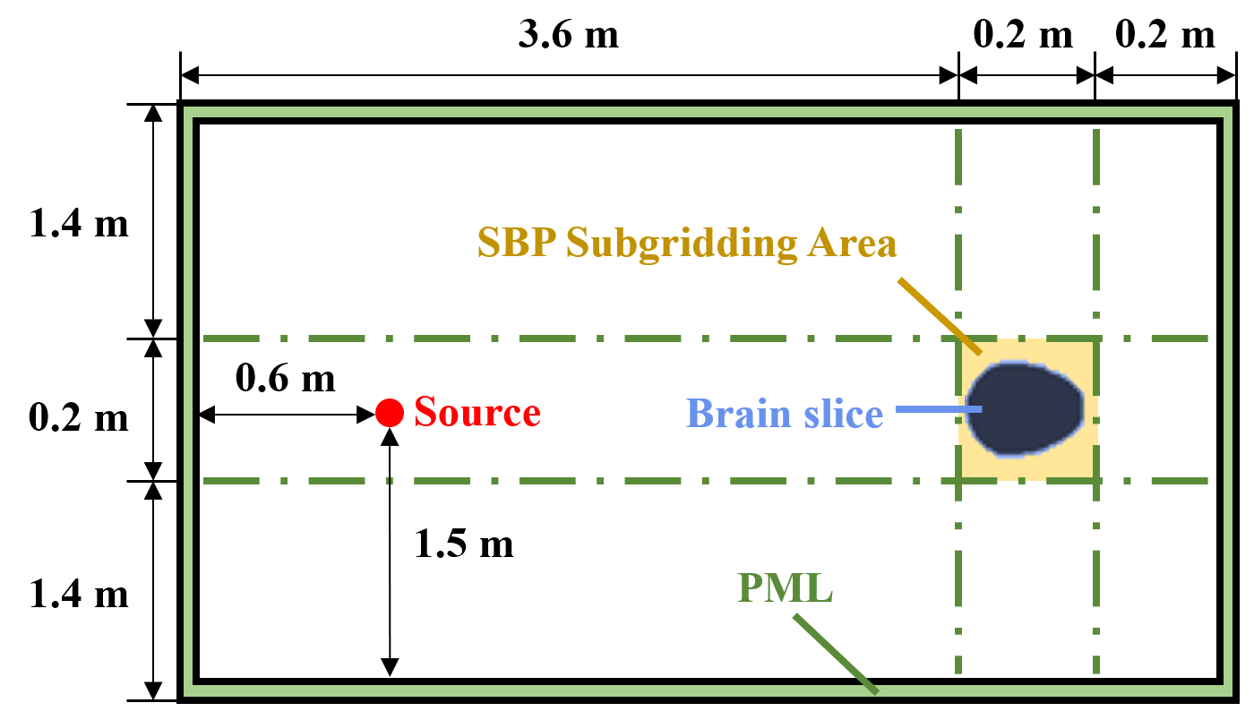}}  	
	\caption{Structure of domain with human head cross-section.}
	\label{figure8}	
\end{figure}

The detailed configurations can be also found in \cite{1}. In our simulation, the detailed parameters of human tissues in Table I are used in our simulation, which is obtained from \cite{1}. The computational domain is terminated with a 10-layer PML, as shown in Fig. \ref{figure8}, according to \cite{29}. The computational domain is $4$ m long and $3$ m wide. The human head is placed at (3.6,1.5) [m], as shown in Fig. \ref{figure8}. The coarse mesh cell size is $4 \times 10^{-3}$ m, and the size of subgridding mesh region is $2 \times 10^{-3}$ m. The simulation time is $1 \times 10^{-7}$ s. A $900$ MHz Gaussian pulse source is placed at (0.6,1.5) [m] in the coarse mesh region. The time step 4.6701 ps is used in all the simulations for fair comparison. According to \cite{14}, the SAR is given by absolute value of the electric field component, the specific conductance and density of the corresponding structure.
\begin{table}
	\centering
	\caption{Detailed parameters of different tissues used in the SAR calculation}\label{table1}
	\begin{tabular}{lccc}
		\hline
		\hline
		\textbf{Tissue} &\textbf{Density [${\bf{kg/{m^3}}}$]  }  & \multicolumn{2}{c}{\textbf{Dielectric Properties}}\cr  
		\hline
		\hline
		&\textbf{Average} &\textbf{$\varepsilon_r$}&\textbf{$\sigma$ [S/m]}\cr
		\hline
		Brain                &   1,046          & 4.0    &   0.04   \\
		\hline
		Cerebrospinal Fluid            &  1,007   & 4.0    & 2.00     \\
		\hline
		Dura             &   1,174        & 4.0      &  0.50      \\
		\hline
		Skull            &   1,908      & 2.5   &  0.02   \cr
		\hline 
		\hline 
	\end{tabular}
\end{table}

\begin{figure}
	\begin{minipage}[h]{0.325\linewidth}\label{ninea}
		\centerline{\includegraphics[scale=0.045]{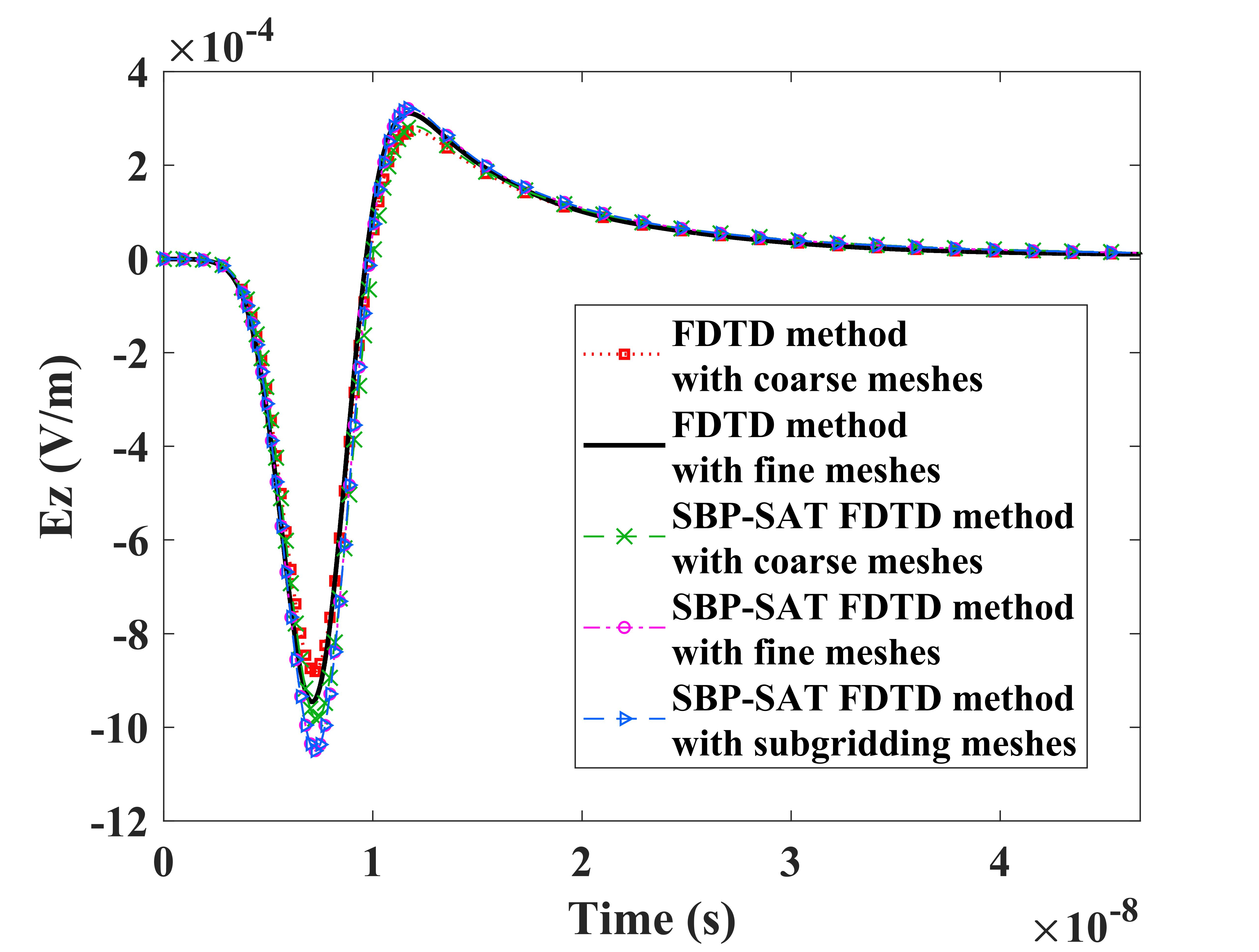}}
		\centerline{(a)}
	\end{minipage}
	\centering \\
	\begin{minipage}[h]{0.325\linewidth}\label{nineb}
		\centerline{\includegraphics[scale=0.045]{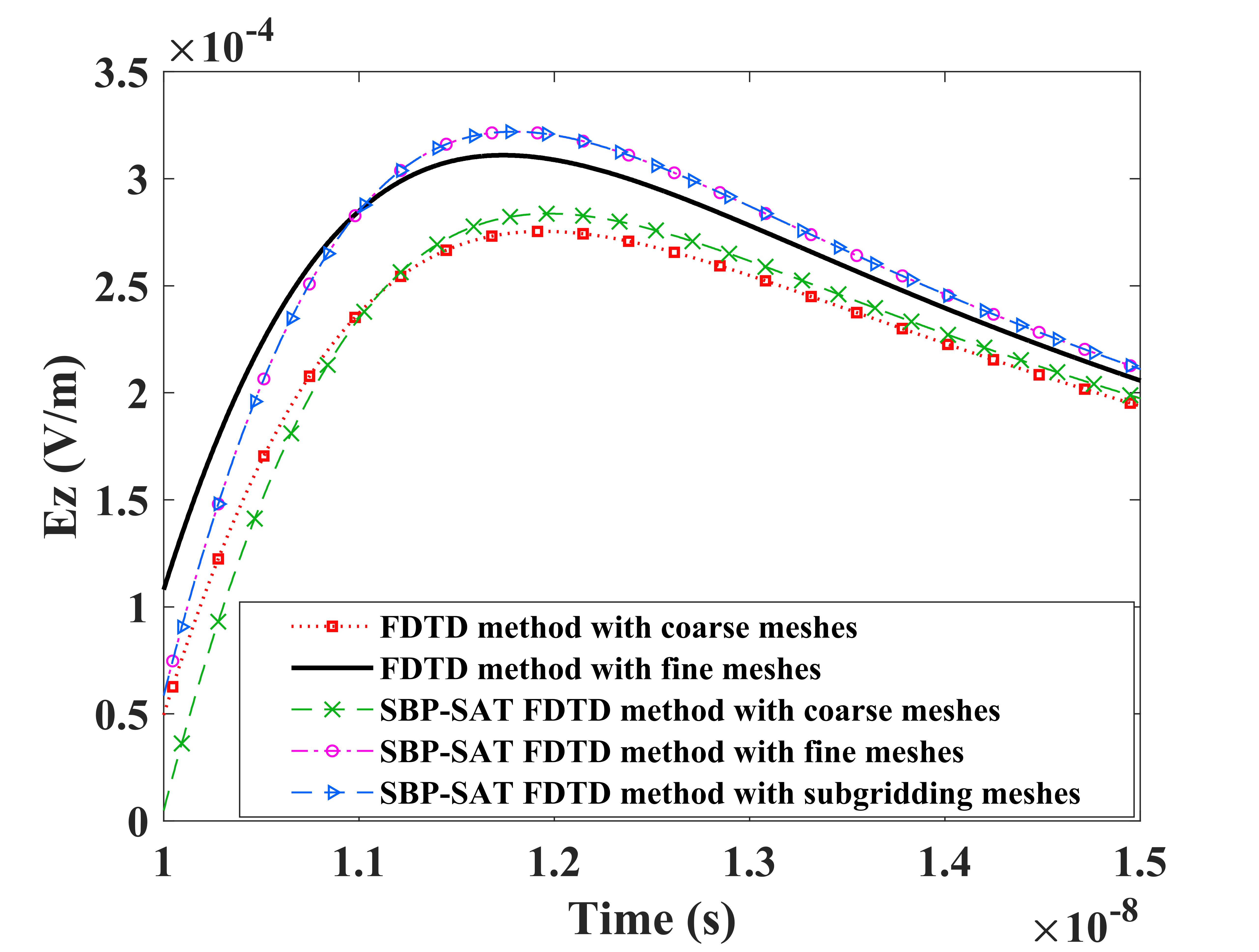}}
		\centerline{(b)}
	\end{minipage}
	\centering
	\caption{ Electric fields in the time domain obtained from the FDTD method and the SBP-SAT FDTD method with different meshes. (a) The waveform in [0,$4.7 \times 10^{-8}$] (s) and (b) [$1.0 \times 10^{-8}$,$1.5 \times 10^{-8}$] (s), respectively.
}
	\label{figure9}
\end{figure}

Fig. \ref{figure9} shows that a Gaussian pulse with the bandwidth of 0.9 GHz is used as the excitation source. The SAR is calculated by the FDTD method, the SBP-SAT FDTD method with different meshes.

\begin{table}
	\centering
	\caption{Computational consumption of the FDTD method and the SBP-SAT FDTD method with different meshes }\label{table2}
	\resizebox{8.77cm}{!}{
		\begin{threeparttable}[b]
			\begin{tabular}{c|c|c|c|c}
				\hline
				\hline
				\textbf{Method} &\textbf{No. of Cells}   & \textbf{Relative Error} & {\textbf{Time Cost}} {[s]}  & {\textbf{Ratio}*}\cr 
				\hline
				\hline
				FDTD fine meshes       &3,000,000 & -         &4845.81  & -    \cr
				\hline
				FDTD coarse meshes      &750,000   & 16.40$\%$ & 1235.72 & 3.92   \\
				\hline
				SBP-SAT FDTD fine meshes &3,000,000 &0.40$\%$   & 5275.71 & 0.92 \\
				\hline
				SBP-SAT FDTD coarse meshes&750,000   &16.90$\%$  & 1345.33 &3.60   \\
				\hline
				SBP-SAT FDTD subgridding meshes&757,500    & 0.61$\%$   & 1414.68 & 3.43  \\
				\hline 
				\hline 
			\end{tabular}
			\begin{tablenotes}
				\footnotesize
				\item[*]Ratio is defined as the ratio of time consumed in the FDTD method with fine meshes to that in the corresponding method.
			\end{tablenotes}
		\end{threeparttable}
	}
\end{table}

Table II shows the computational consumption of the relative error, the total number of cells and time cost of different methods. There are $7.5 \times 10^{5}$ cells in the coarse grid and $3 \times 10^{6}$ cells in the fine grid. In the subgridding meshes, the number of cells ($7.5 \times 10^{5}$) slightly increased. As shown in Table II, the FDTD method and the SBP-SAT FDTD method with fine meshes can obtain accurate results. The CPU time of the two methods is 4845.81 s and 5275.71 s, respectively. Compared with the methods with coarse meshes, they use more CPU time to complete the simulation. It is worth noting that the SBP-SAT FDTD method uses more CPU time than the FDTD method due to the cost of calculating SAT items. However, for the proposed subgridding method, since the fine grid is only used in small regions containing the human head model, the total number of cells is $7.575 \times 10^{5}$, which is only 1.0\% more than that of the coarse grid. At the same time, the relative error of the proposed SBP-SAT FDTD method is only 0.61\%, which shows significant accuracy improvement.

\subsection{The Substrate Integrated Waveguide (SIW)} 
\begin{figure}
	\centerline{\includegraphics[scale=0.35]{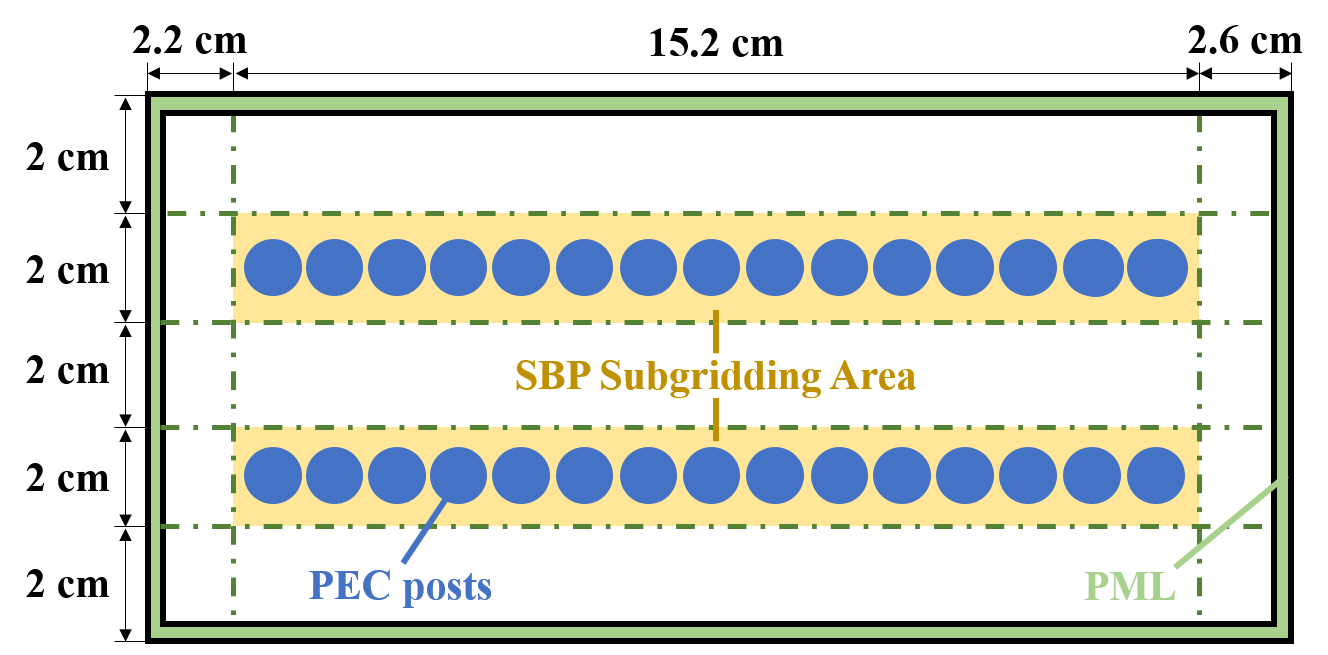}}  	
	\caption{Geometrical Configurations of the SIW and the computational domain distributions.}
	\label{figure10}	
\end{figure}

Finally, a substrate integrated waveguide (SIW) with two rows of PEC posts on the sidewall is simulated, as shown in Fig. \ref{figure10}. The radius of each post is $4$ mm. Fig. \ref{figure10} shows the discretization of the SIW structures. The entire domain is divided into 15 regions, and the SBP-SAT FDTD subgridding method is used to simulate the region including PEC posts. In the other regions, coarse meshes are used in the SBP-SAT FDTD method.

\begin{figure}
	\begin{minipage}[h]{0.325\linewidth}\label{elevena}
		\centerline{\includegraphics[scale=0.05]{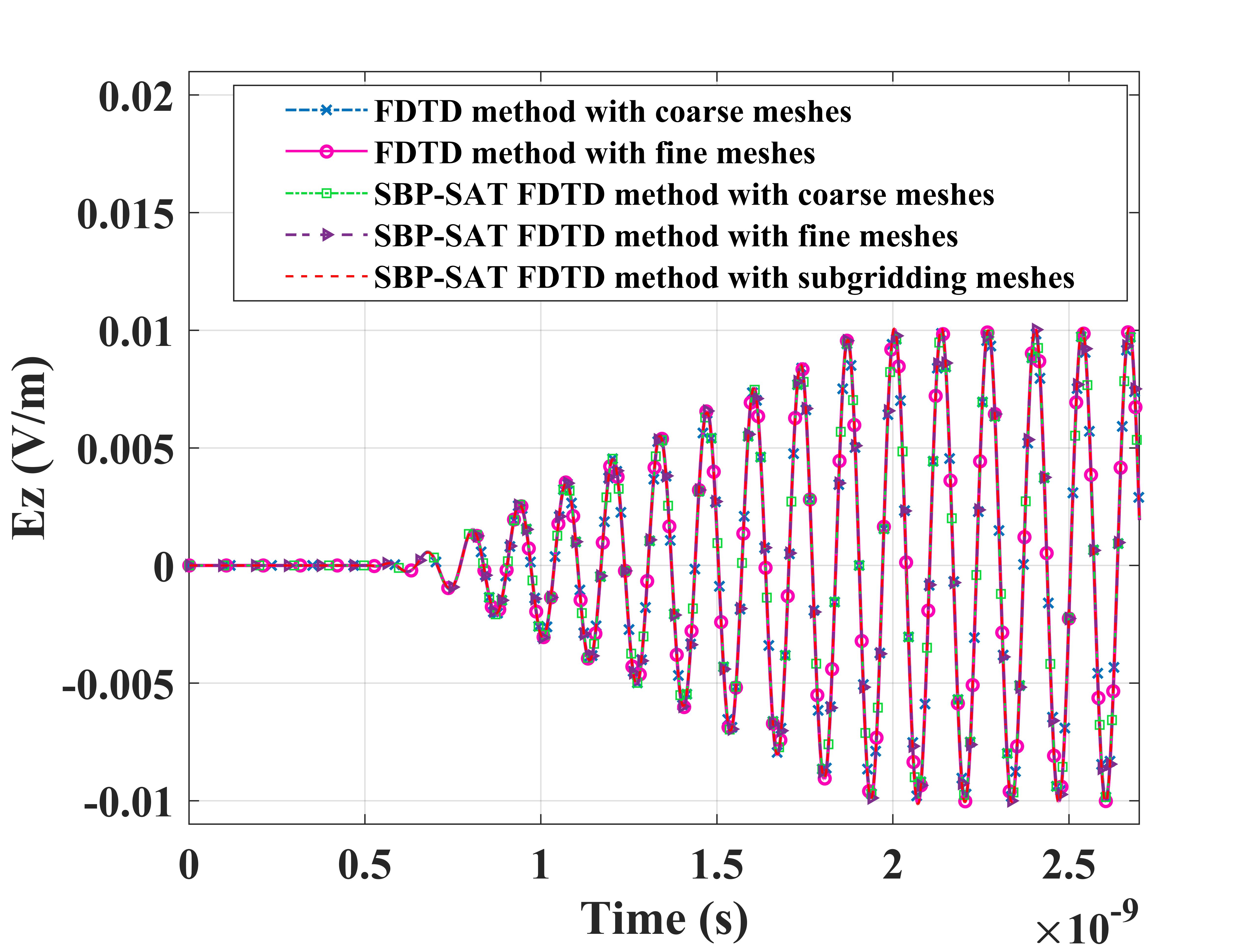}}
		\centerline{(a)}
	\end{minipage}
	\centering \\
	\begin{minipage}[h]{0.325\linewidth}\label{elevenb}
		\centerline{\includegraphics[scale=0.049]{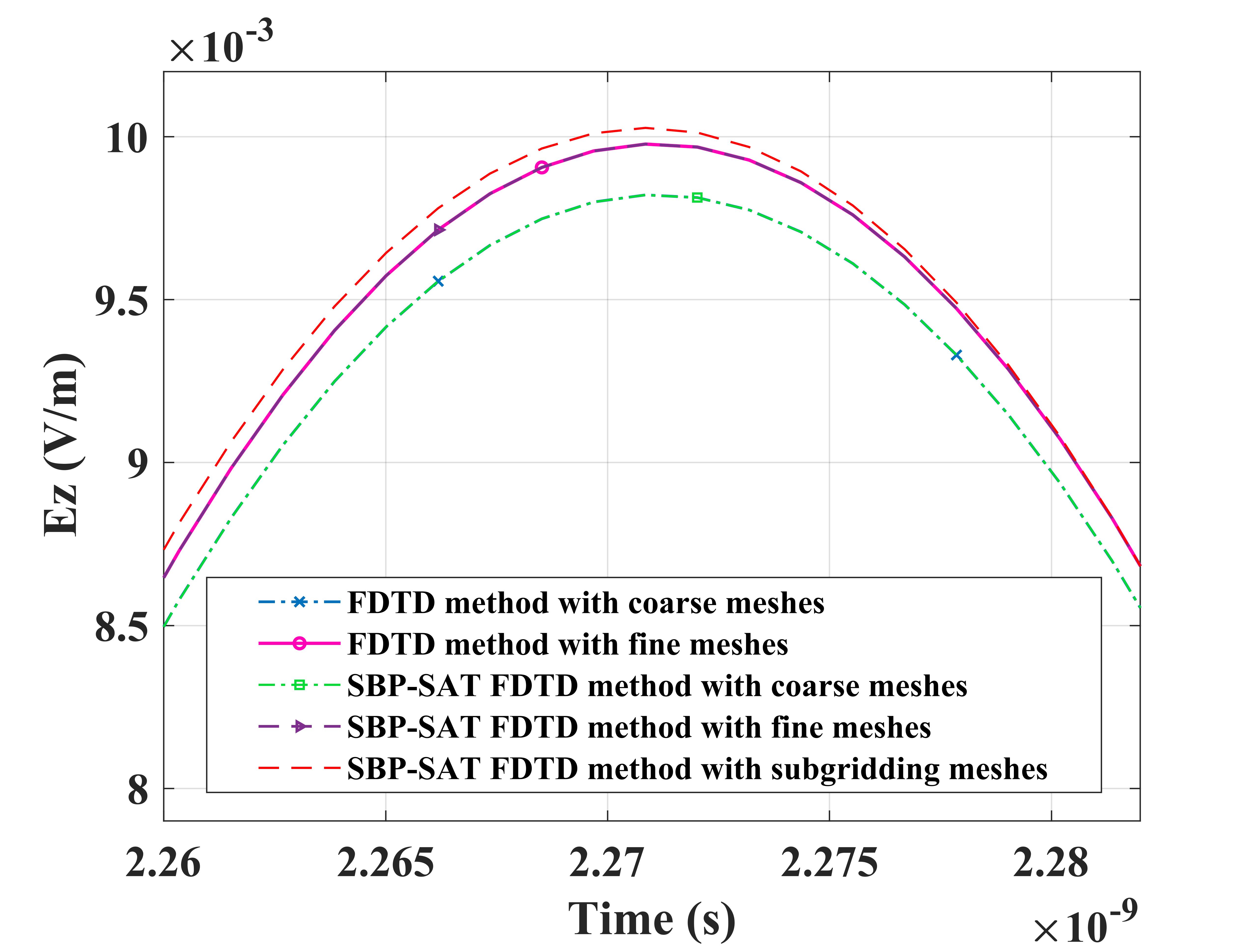}}
		\centerline{(b)}
	\end{minipage}
	\centering
	\caption{Electric fields in the time domain obtained from the FDTD method and the SBP-SAT FDTD method with different meshes. (a) The waveform in [0,$2.650 \times 10^{-9}$] (s) and (b) [$2.260 \times 10^{-9}$,$2.282 \times 10^{-9}$] (s), respectively.}
	\label{figure11}
\end{figure}
 
The cell size of coarse meshes is $1$ mm and $0.5$ mm for subgridding meshes. The time step is $0.99$ times that of the CFL condition in the subgridding region. The current source is used to excite the EM field. The ramp sinusoidal function of $\text{TE}_{10}$-mode is $7.5$ GHz.
The simulation time is $8 \times 10^{-9}$ s. A $7.5$ GHz sinusoidal source is placed at (17,5) [cm] in the coarse mesh region. The time step 1.1675 ps is used in all the simulations for fair comparison.

Fig. \ref{figure11} shows the comparison of results obtained from  different methods in the time domain. It can be found that the results of different methods are very similar. However, the accuracy of the proposed SBP-SAT FDTD method with subgridding meshes is similar to that of the FDTD and the SBP-SAT FDTD method with fine meshes, and is better than that of the FDTD and SBP-SAT FDTD method with coarse grids.

\begin{figure} 	
	\begin{minipage}[h]{0.48\linewidth}\label{figure12a}
	\centerline{\includegraphics[scale=0.067]{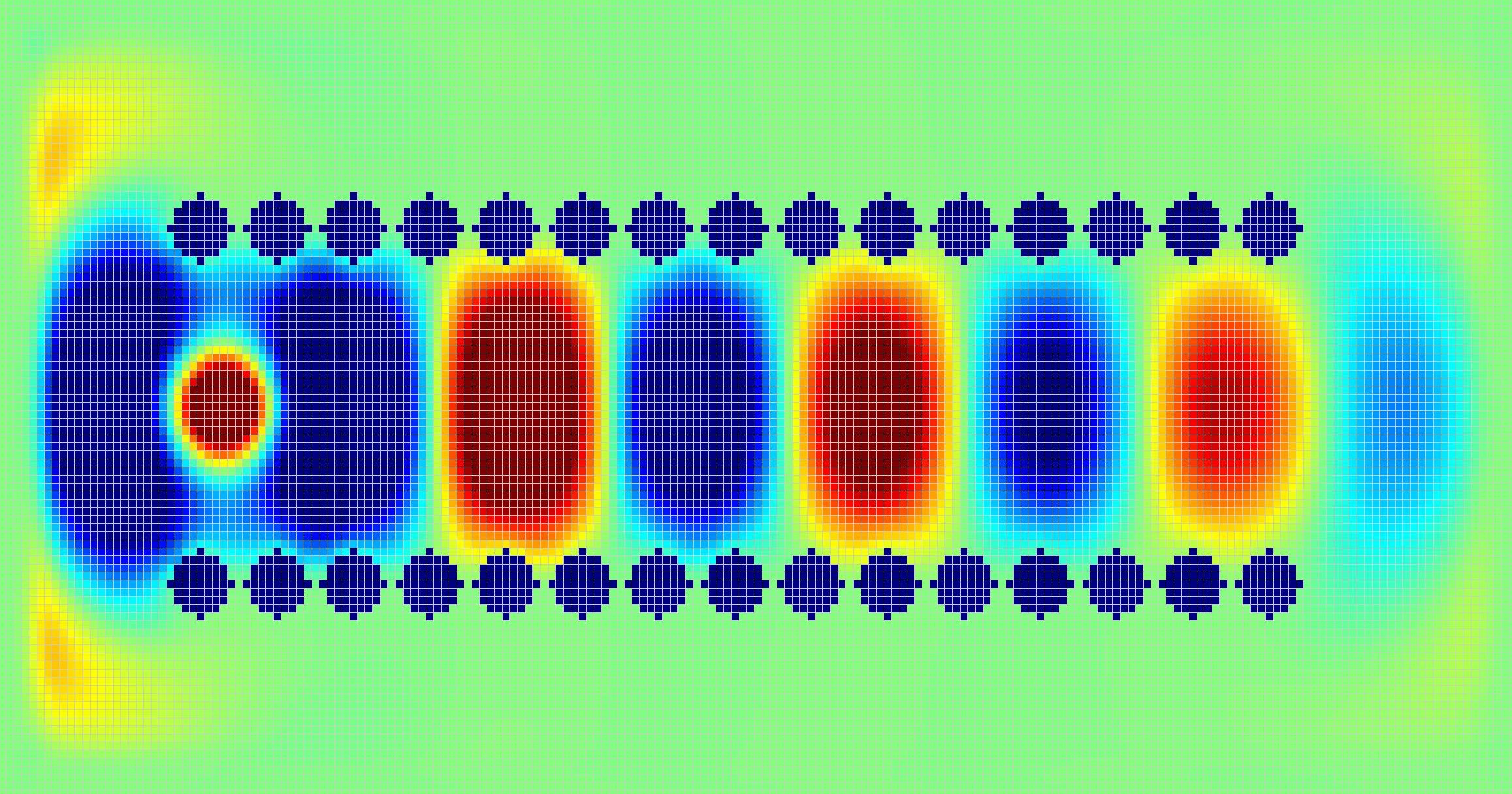}}
	\centerline{(a)}
\end{minipage}
\centering
\begin{minipage}[h]{0.48\linewidth}\label{figure12b}
	\centerline{\includegraphics[scale=0.033]{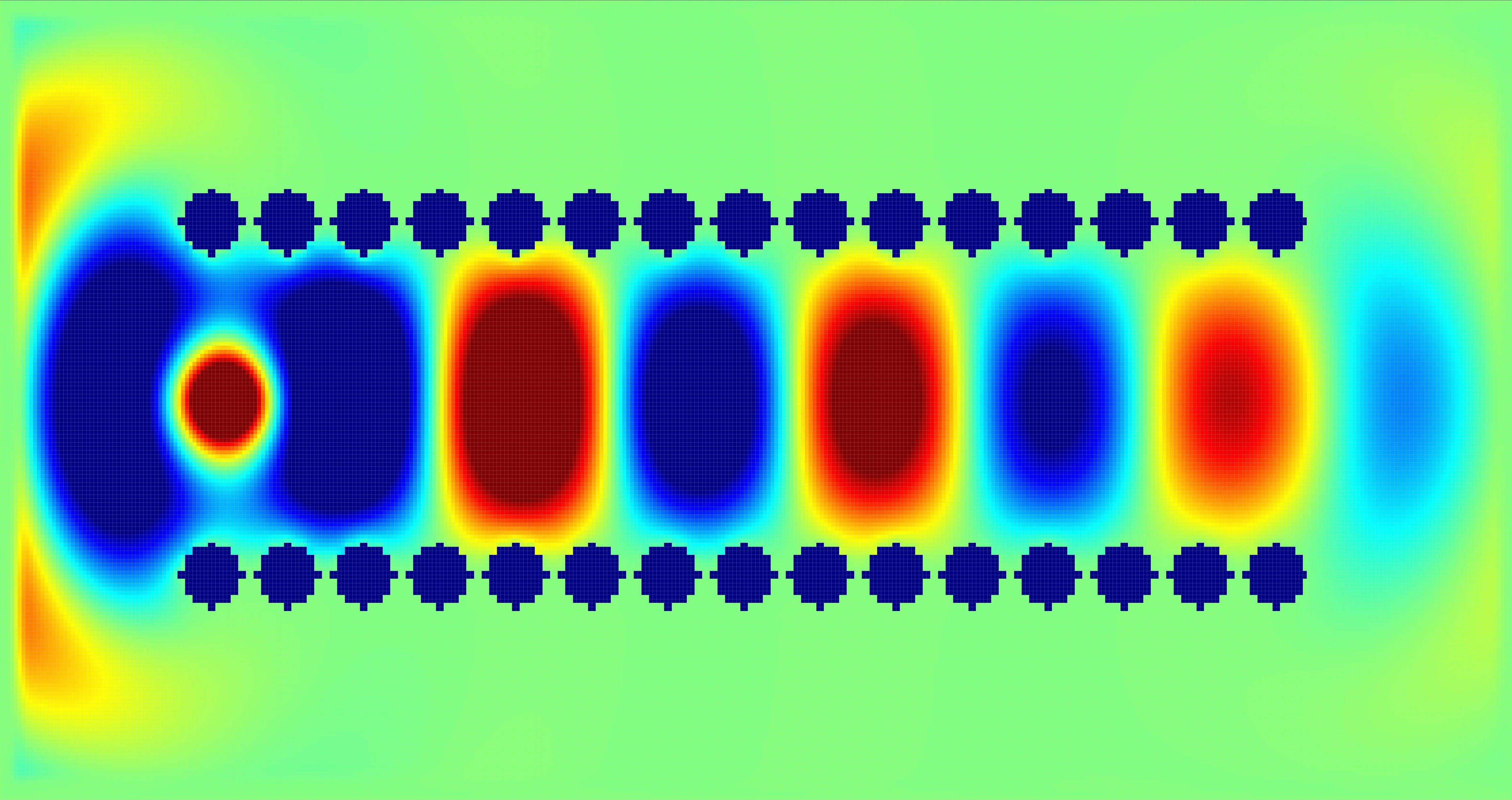}}
	\centerline{(b)}
\end{minipage}
	\centering \\
	\begin{minipage}[h]{0.48\linewidth}\label{figure12c}
	\centerline{\includegraphics[scale=0.067]{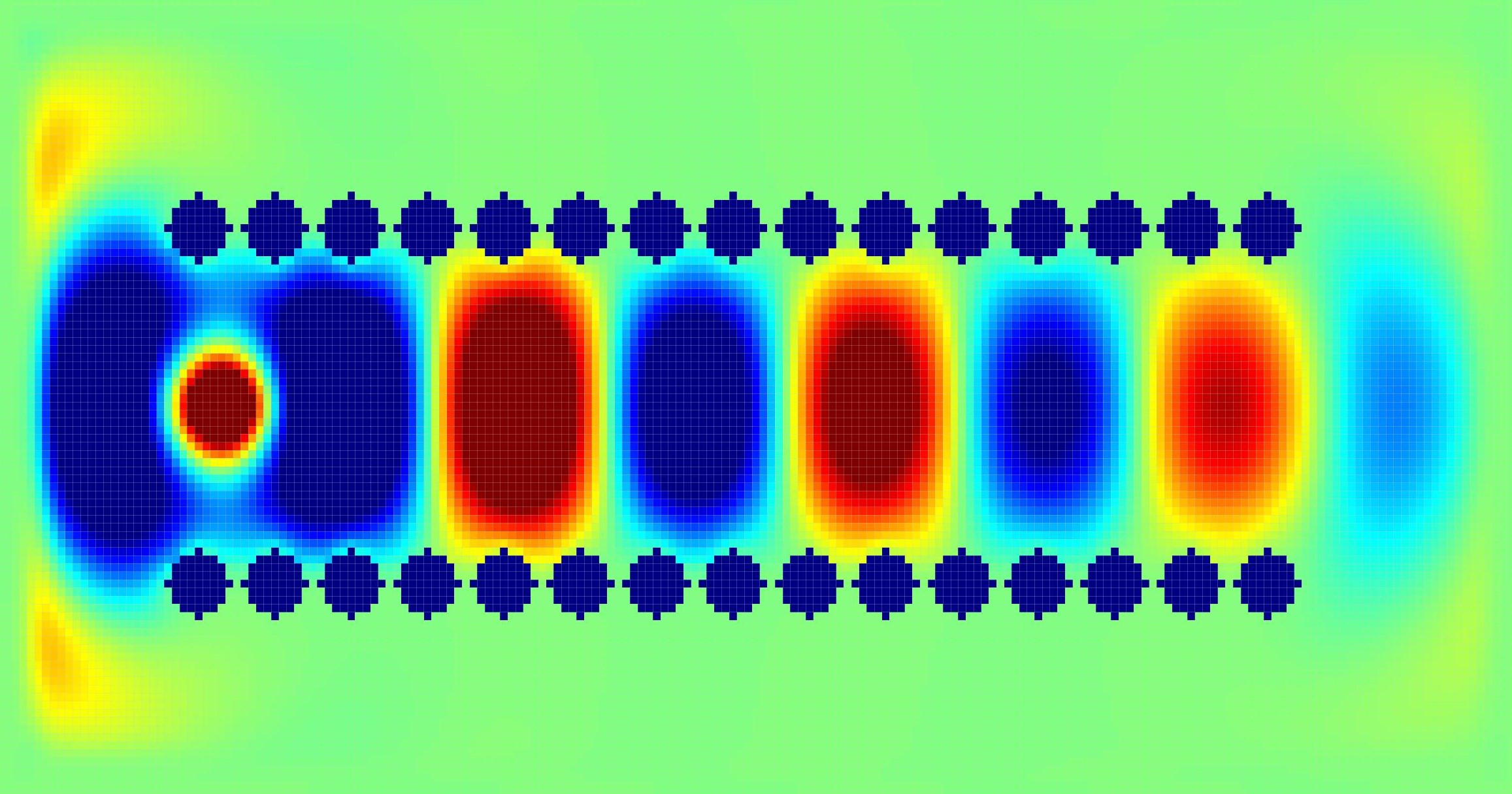}}
	\centerline{(c)}
\end{minipage}
\centering
\begin{minipage}[h]{0.48\linewidth}\label{figure12d}
	\centerline{\includegraphics[scale=0.033]{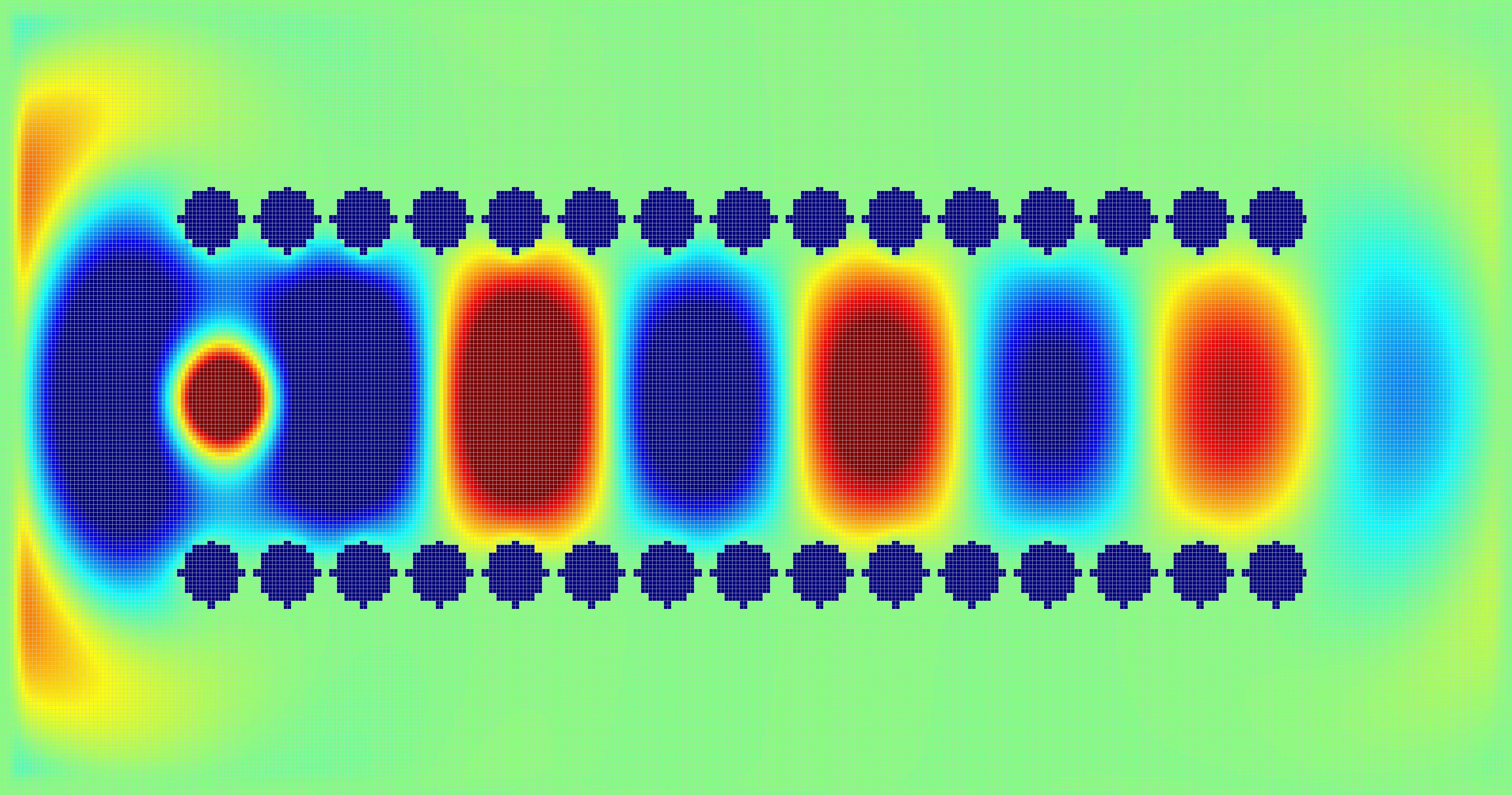}}
	\centerline{(d)}
\end{minipage}
\centering \\
	\begin{minipage}[h]{0.05\linewidth}\label{figure12e}
		\centerline{\includegraphics[scale=0.068]{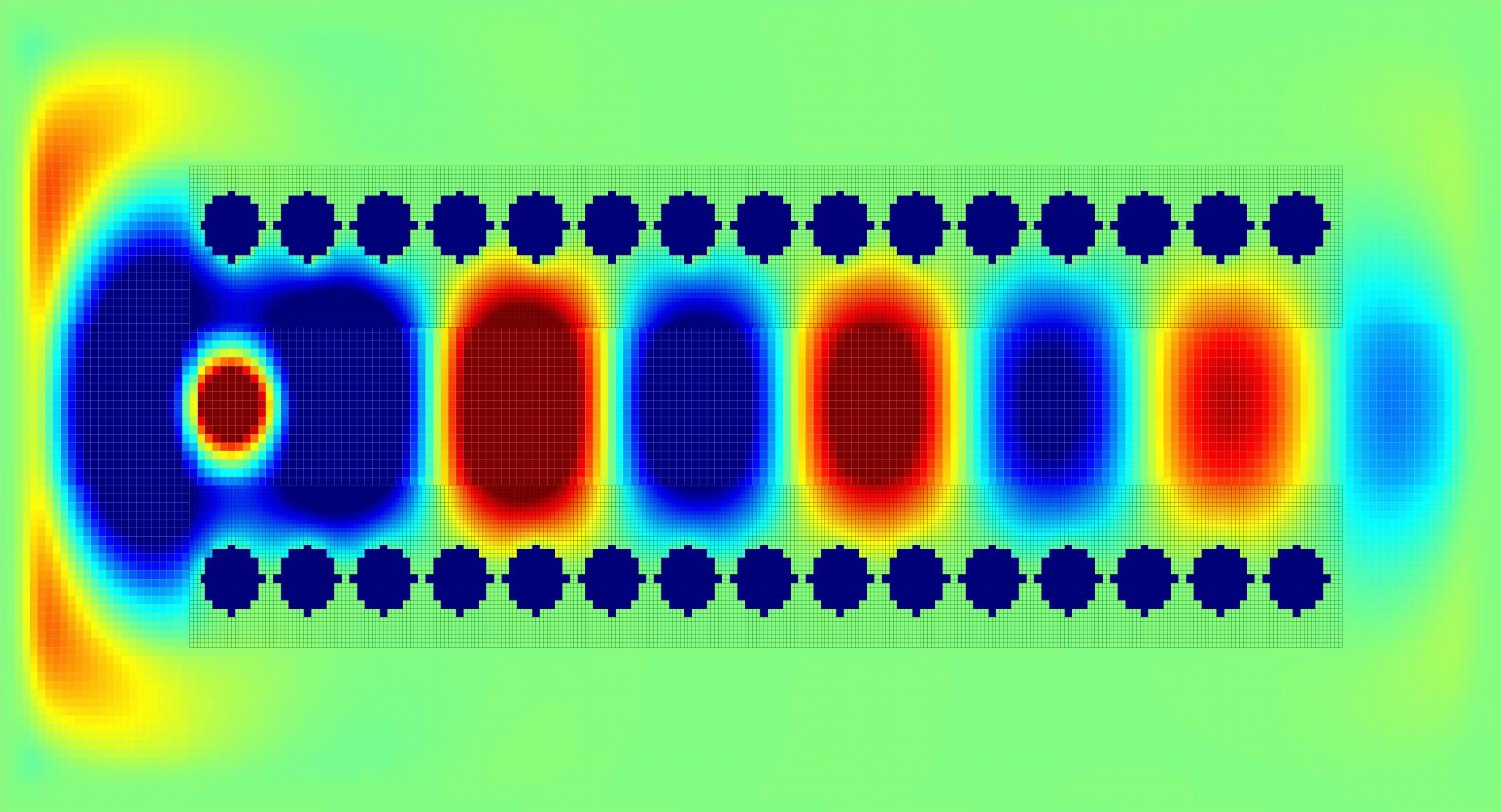}}
		\centerline{(e)}
	\end{minipage}
	\centering
	\caption{$\text{TE}_{10}$-mode obtained with the different methods. (a) the FDTD method with coarse meshes, (b) the FDTD method with fine meshes, (c) the SBP-SAT FDTD method with coarse meshes, (d) the SBP-SAT FDTD method with fine meshes, (e) the proposed SBP-SAT FDTD method with subgridding meshes. }
	\label{figure12}
\end{figure}

We compare the $\text{TE}_{10}$-mode in the SIW region obtained by the FDTD method, the SBP-SAT FDTD method and the proposed SBP-SAT FDTD method with different meshes, as shown in Fig. \ref{figure12}. In Fig. \ref{figure12}(a) and \ref{figure12}(c), since coarse meshes are used in the FDTD method and the SBP-SAT FDTD method, results are very similar. In Fig. \ref{figure12}(b) and \ref{figure12}(d), due to fine meshes in the two methods, more details are shown. Fig. \ref{figure12}(e) shows the $\text{TE}_{10}$-mode calculated by the proposed SBP-SAT FDTD subgridding method. Fine grids are used in two regions which cover PEC posts, so there are no visible differences in the PEC post regions in Fig. \ref{figure12}(e) and Fig. \ref{figure12}(b) and \ref{figure12}(d).

\begin{table}
	\centering
	\caption{Computational consumption of the FDTD method and the SBP-SAT FDTD method with different meshes}\label{table2}
	\resizebox{8.77cm}{!}{
		\begin{threeparttable}[b]
			\begin{tabular}{c|c|c|c|c}
				\hline
				\hline
				\textbf{Method} &\textbf{No. of Cells}   & \textbf{Relative Error} & {\textbf{Time Cost}} {[s]}  & {\textbf{Ratio}*}\cr 
				\hline
				\hline
				FDTD fine meshes       &8,000,000 & -         &183.32  & -    \cr
				\hline
				FDTD coarse meshes      &2,000,000   & 2.03$\%$ & 49.28 & 3.72   \\
				\hline
				SBP-SAT FDTD fine meshes &8,000,000 & 0.00$\%$   & 211.98 & 0.86 \\
				\hline
				SBP-SAT FDTD coarse meshes&2,000,000   & 2.03$\%$  & 51.20 &  3.58 \\
				\hline
				SBP-SAT FDTD subgridding meshes&3,824,000    & 0.57$\%$ & 63.06 & 2.91  \\
				\hline 
				\hline 
			\end{tabular}
			\begin{tablenotes}
				\footnotesize
				\item[*]Ratio is defined as the ratio of time consumed in the FDTD method with fine meshes to that in the corresponding method.
			\end{tablenotes}
		\end{threeparttable}
	}
\end{table}

The computational time and efficiency are also carefully exampled. Table III shows the calculation consumption of relative error, the total number of cells and time cost by different methods. There are $2 \times 10^{6}$ cells in the coarse grid and $8 \times 10^{6}$ cells in the fine grid. In the subgrid, the number of cells ($3.824 \times 10^{6}$) increased slightly. As shown in table III, the FDTD method and the SBP-SAT FDTD method with fine meshes can obtain accurate results. CPU time of the two methods is 183.32 s and 211.98 s, respectively. For the proposed subgridding method, since the fine grid is only used in small regions containing the PEC posts, the total number of cells is $3.824 \times 10^{6}$. However, the relative error of the proposed method is only 0.57\%.


\section{Conclusion}  
In this paper, we proposed an SBP-SAT FDTD subgridding method to model the geometrically fine structures. Through field extrapolation on the boundaries to make the discrete operators satisfy the SBP property, the proposed SBP-SAT FDTD method uses the Yee’s grid without adding or modifying any field components. In addition, SATs is used to weakly enforce tangential boundary conditions between different grid blocks. By carefully designing the interpolation matrix and choosing the free parameters of SAT, the whole computational domain has no dissipation. Therefore, its long-term stability is theoretically guaranteed. Numerical results validate its stability, accuracy and efficiency. It shows great potential in practical engineering applications.

%
%

\end{document}